# THE LOCAL GROUP OF GALAXIES*

**Sidney van den Bergh**
National Research Council of Canada
Dominion Astrophysical Observatory
Herzberg Institute of Astrophysics
5071 West Saanich Road
Victoria, British Columbia, V8X 4M6
Canada
e-mail: vdb@dao.nrc.ca

## Summary

Hubble's (1936, p. 125) view that the Local Group (LG) is "a typical, small group of nebulae which is isolated in the general field" is confirmed by modern data. The total number of certain and probable Group members presently stands at 35. The half-mass radius of the Local Group is found to be $R_h \approx 350$ kpc. The zero-velocity surface, which separates the Local Group from the field that is expanding with the Hubble flow, has a radius $R_o = 1.18 \pm 0.15$ Mpc. The total mass of the LG is $M_{LG} = (2.3 \pm 0.6) \times 10^{12} M_\odot$. Most of this mass appears to be concentrated in the Andromeda and Milky Way subgroups of the LG. The total luminosity of the Local Group is found to be $M_V = -22.0$: This yields a mass-to-light ratio (in solar units) of $M/L_V = 44 \pm 12$. The solar motion with respect to the LG is $306 \pm 18$ km s$^{-1}$, directed towards an apex at $\ell = 99° \pm 5°$, and $b = -4° \pm 4°$. The velocity dispersion within the LG is $\sigma_r = 61 \pm 8$ km s$^{-1}$. The galaxies NGC 3109, Antlia, Sextans A and Sextans B appear to form a distinct grouping with $V_r = +114 \pm 12$ kpc relative to the LG, that is located

---

beyond the LG zero-velocity surface at a distance of 1.7 Mpc from the Local Group centroid.  The luminosity distribution of the LG has a slope $\alpha$ = -1.1 $\pm$ 0.1. This value is significantly less negative than that which is found in rich clusters of galaxies.  The luminosity distribution of the dwarf spheroidal galaxies is steeper than that for dwarf irregulars.  Furthermore the dSph galaxies are strongly concentrated within the Andromeda and Milky Way subclusters of the Local Group, whereas the majority of dIr galaxies appear to be free-floating members of the LG as a whole.  With the possible exception of Leo I and Leo A, most LG members appear to have started forming stars simultaneously  ~ 15 Gyr ago. Many of the galaxies, for which evolutionary data are available, appear to have shrunk with time.  This result is unexpected because Hubble Space Telescope observations appear to show galaxies at z ~ 3 to be smaller than they are at z = 0. In the Large Magellanic Cloud the rate of cluster formation was low for a period that extended from  ~ 12 Gyr  to  ~ 4 Gyr ago.  The rate of cluster formation may have increased more rapidly 3-5 Gyr ago, than did the rate of star formation.  The reason for the sudden burst of cluster formation in the LMC  ~ 4 Gyr ago remains obscure.  None of the dwarf galaxies in the LG appears to have experienced a starburst strong enough to have produced a "boojum".

1.      **Introduction**

Hubble (1936) was the first to draw attention to the fact that our Milky Way system belongs to a small cluster of galaxies, which he dubbed The Local Group.  Inspection of the prints of the <u>Palomar Sky Survey</u> shows that about half of all galaxies in the Universe are situated in such small clusters.  In other words the Galaxy is located in a rather typical region of the Universe.  Hubble emphasized that the Local Group was important since it (1) provided an opportunity to study nearby examples of a wide range of galaxy types in detail, and (2) because it enabled astronomers to calibrate the luminosities of "standard candles" such as Cepheids, RR Lyrae stars etc., which could then be used to determine the extragalactic distance scale.  For modern reviews on the galaxies of the Local Group the reader is referred to Grebel (1997) and Mateo (1998), to <u>IAU Symposium No. 192</u> (Whitelock and Cannon 1999), and to the monograph <u>The Galaxies of the Local Group</u> (van den Bergh 2000).

2.      **Local Group membership**

In <u>The Realm of the Nebulae</u> Hubble assigned the following galaxies to the Local Group:  The Milky Way system (= the Galaxy) and its companions the Large Magellanic Cloud (LMC) and the Small Magellanic Cloud (SMC), the Andromeda Galaxy (M 31 = NGC 224), and its companions M 32 (NGC 221) and NGC 205, the Triangulum Galaxy (M 33 = NGC 598), NGC 6822 and IC 1613.  Of three additional Local Group (LG) suspects listed by Hubble only one (IC 10) is, in fact, close enough to be regarded as a <u>bona fide</u> LG member.  The number of known LG members was increased from 10 to 19 by Baade (1963), who added the late-type dwarf galaxies NGC 147, NGC 185, Sculptor, Fornax, Leo I (Regulus), Leo II, Fornax, Draco and Ursa Minor to the roster of LG members.  In recent years the number of known LG members has, on average, grown by four or five per decade.  The number of LG members and suspected members increases by the discovery of new faint members, or by the recognition of more luminous objects in rich star fields (or behind dense absorbing clouds) in the plane of the Milky Way.  On the other hand the number of suspected members has decreased by the elimination of objects that turn out to be associated with the nearby Antlia-Sextans (van den Bergh 1999a), IC 342/Maffei and Sculptor (South Polar) groups. Improved distance estimates for individual galaxies have added  (and subtracted) from the list of likely LG members.  A compilation of data on certain and probable LG members, which is based on van den Bergh (2000), is given in Table 1.  This table gives positions, morphological types, distances of these objects D

Place Table 1 here

from the Sun, and $D_{LG}$ from the barycenter of the Local Group, which was assumed to be located at 0.6 times the distance to M 31, at D =  454 kpc towards $\ell = 121°.7$ and b = -21°.3.

### 3. The nearest neighbors of the Local Group

The Local Group is one of many small groups and clusters of galaxies that are situated along the outer edge of the Virgo supercluster. The nearest neighbors of the LG are the Antlia-Sextans group at $D_{LG}$ = 1.7 Mpc (van den Bergh 1999a), the Sculptor group at $D_{LG} \approx$ 2.4 Mpc (Jergen, Freeman & Binggeli 1998), the IC 342/Maffei group at $D_{LG}$ = 3.2 (Krismer, Tully & Gioia 1995) and the M 81 group at $D_{LG}$ = 3.5 Mpc (Freedman et al. 1994). Because of their proximity the Ant-Sex group galaxies NGC 3109, Antlia, Sextans A and Sextans B have been regarded as members of the LG by some authors. However, the fact that (1) all four of these galaxies lie well beyond the LG zero-velocity surface at $R_o$ = 1.18 ± 0.15 Mpc, and (2) that they are redshifted by $<V_r> = +114 \pm 12$ km s$^{-1}$ suggests (van den Bergh 1999a) that these objects, in fact, constitute the nearest small external group of galaxies. A search for very faint dIr or dSph galaxies in Antlia, Hydra and Sextans might prove rewarding.

### 4. Spatial distribution of Local Group members

A plot of the spatial distribution of LG members (Courteau & van den Bergh 1999) is given in Figure 1. This figure shows that the Local Group is a

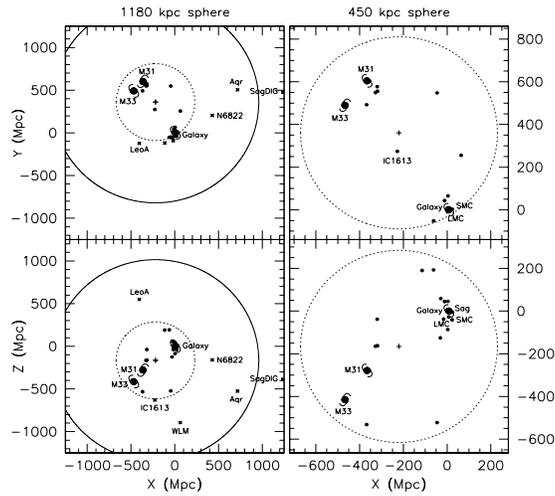

**Figure 1:** Distribution of Local Group members, as viewed from two orthogonal directions. The X, Y and Z axes point towards the Galactic center ($\ell = 0°$, b $= 0°$), the direction of rotation ($\ell = 90°$, b $= 0°$) and the north Galactic pole (b $= +90°$), respectively. The figure shows that the majority of LG members are concentrated in two subgroups that are centered on the Galaxy and M 31. The circle of radius of 1180 kpc corresponds to the zero-velocity surface of the LG. The dashed circle with radius 450 kpc shows the radius of the sphere that contains half of all Local Group galaxies.

cluster with a binary core. The majority of Local Group members are seen to belong to either the Andromeda subgroup, or to the Galactic subgroup, of the LG. The radial distribution of LG members is shown in Figure 2 (differential) and

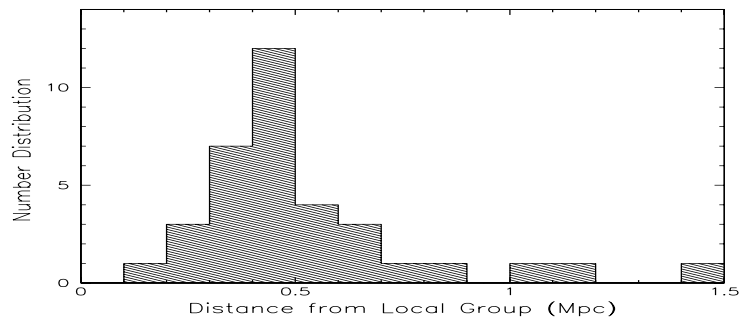

**Figure 2:** Distribution of distances of individual LG members from the adopted barycenter. Only three LG suspects are seen to have $D_{LG} > 1.0$ Mpc.

Figure 3 (integral). These figures show that the LG is a centrally concentrated

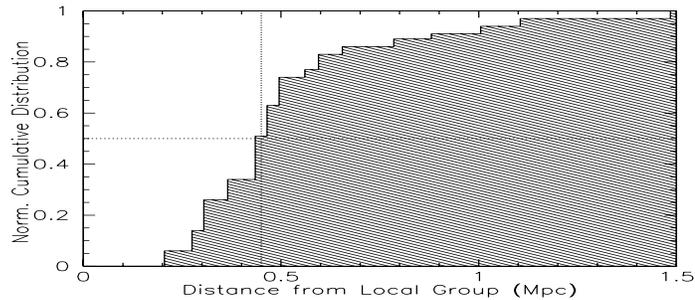

**Figure 3:** Integral distance distribution of LG members. Half of all Local Group galaxies are seen to be located within ~ 450 kpc of the adopted barycenter

cluster, in which half of all members lie within ~ 450 kpc of the adopted

barycenter. (Because of the high masses of M 31 and the Galaxy the half-mass

radius of the Local Group is estimated to be only ~ 350 kpc). The low galaxy

density near $D_{LG} = 0$ is, of course, due to the fact that the barycenter of the LG

lies between the M 31 and Galaxy subgroups.

Table 2 shows that the distribution of luminous LG galaxies with $M_V < -$

Place Table 2 here

14.0 exhibits a hierarchical structure. In the Andromeda subgroup M 31 + M 32 +

NGC 205 form a physical triple system, and NGC 147 + NGC 185 constitute a

physical binary (van den Bergh 1998c). Within the Milky Way subgroup the

LMC + SMC form a physical double. It is of interest to note that all three of the galaxies (NGC 6822, IC 1316, WLM) in Table 2, that are not members of subgroups, are irregulars. Among the 14 LG galaxies with -14.0 < $M_V$ < -10.0 it is also found that the dwarf irregular galaxies (DDO 216 = Pegasus), Leo A and SagDIG are free-floating members of the LG, whereas eight or nine dSph galaxies (Sagittarius, Fornax, Leo II, And I, And II, And III, And V, And VI, and perhaps Leo I) are members of either the Andromeda or Galactic subgroups of the LG. This result (van den Bergh 1999b), which was first noted by Einasto et al. (1974), suggests that dSph galaxies primarily form deep in the potential wells of massive galaxies, whereas irregulars are mostly assembled far from major galaxies. The only presently known exception to this rule is provided by the faint, remote, and isolated, Tucana dwarf spheroidal. Van Driel et al. (1998) have noted a similar phenomenon among the faint galaxies in the nearby M 81 group. They find that the dSph galaxies are mainly concentrated in the core of this group, whereas the dIr galaxies are mostly distributed in the outer reaches of the M 81 group. It is now known that dwarf spheroidal galaxies exhibit a wide variety of evolutionary histories (cf. Grebel 1997, Mateo 1998). It would be of great interest (van den Bergh 1994a) to know if dSph galaxies such as Draco and Ursa Minor, in which almost all stars were formed long ago, are distributed differently than objects like Leo I and Carina, in which the majority of stars are only a few Gyr old. Some possible evidence in favor of this view is provided by the observation that the Pisces system (= LGS 3), that is still forming a few stars at the present time, appears to lie near the outer edge of the Andromeda subgroup of the Local Group.

The Local Group contains five irregular galaxies brighter than $M_V = -14$. Among these NGC 6822 and IC 1613 are not members of subgroups. Although IC 10 appears projected on the outer part of the Andromeda subgroup it may not be a physical member of it. Very recently Sakai, Madore & Freedman (1999) have obtained distances of 500 - 660 kpc for this galaxy. Such a distance would place IC 10 slightly in front of the Andromeda subgroup of the Local Group, which is centered at a distance of 760 kpc. This suggests the possibility that the irregular galaxy IC 10 might, like NGC 6822 and IC 1613, be a free-floating member of the Local Group. The Magellanic Clouds would appear to be an exception to the rule that irregular galaxies are not formed within subgroups of the Local Group. However, orbital simulations by Byrd et al. (1994) suggest that the LMC and SMC might have formed in the neighborhood of M 31 ~ 10 Gyr ago, and that they were subsequently captured by the Galaxy ~ 6 Gyr ago. It should, however, be emphasized that such computations are uncertain and involve a large number of free parameters. In summary, it still appears premature to conclude with certainty that all Ir galaxies are formed outside the dense subclusters within the Local Group.

It is of interest to note that the only two LG binary systems with comparable masses also have similar morphological types. The LMC and SMC are both irregular galaxies, whereas NGC 147 and NGC 185 are both spheroidals. This result suggests that galaxies of comparable masses, that form in the same kinds of environments, evolve into similar types of galaxies. It is also noted in passing that the three most (massive) LG members (M 31, the Galaxy and M 33) are all spiral galaxies.

## 5. Mass of the Local Group

Figure 4 shows a plot of the radial velocities versus the cosines of the apex distances for those LG members for which radial velocity observations are available. Courteau & van den Bergh (1999) have used these data to derive a solar velocity V = 306 ± 18 km s$^{-1}$, towards $\ell$ = 99° ± 5° and b = -4° ± 4°. The quoted errors were determined by using "bootstrap" techniques (Diaconis & Efron 1983).  The dispersion around the regression line plotted in Figure 4 is $\sigma_r = 61 \pm 8$ km s$^{-1}$.  Substituting this number, together with the value $R_h \approx 350$ kpc that was obtained above, into the relation (Spitzer 1969, Binney & Tremaine 1987, Eq. 4-80b)

$$M_{LG} \approx 7.5 <\sigma_r^2> R_h/G = 1.74 \text{ x } 10^6 <\sigma_r^2> R_h \, M_\odot \, , \qquad (1)$$

in which $R_h$ is in kpc, and $\sigma_r$ in km s$^{-1}$, gives $M_{LG} = (2.3 \pm 0.6)$ x $10^{12} \, M_\odot$ .  It is of historical interest to note how close this value is to $M_{LG} \gtrsim 1.8$ x $10^{12} \, M_\odot$ , that Kahn & Woltjer (1959) derived from the approach velocity between M 31 and the Galaxy, and an assumed LG age $T_o = 15$ Gyr.  With the LG mass derived above, and an assumed LG age of $14 \pm 2$ Gyr, the radius of the zero-velocity surface of the LG, i.e. the radius beyond which galaxies participate in the Hubble expansion of the Universe, is (Lynden-Bell 1981, Sandage 1986)

$$R_o \text{ [Mpc]} = (8GT_o^2 \, M_{LG}/\pi^2)^{\frac{1}{3}} = 0.154(T_o \text{ [Gyr]})^{\frac{2}{3}} \, (M_{LG} \text{ [}10^{12} \, M_\odot\text{]})^{\frac{1}{3}} \, . \quad (2)$$

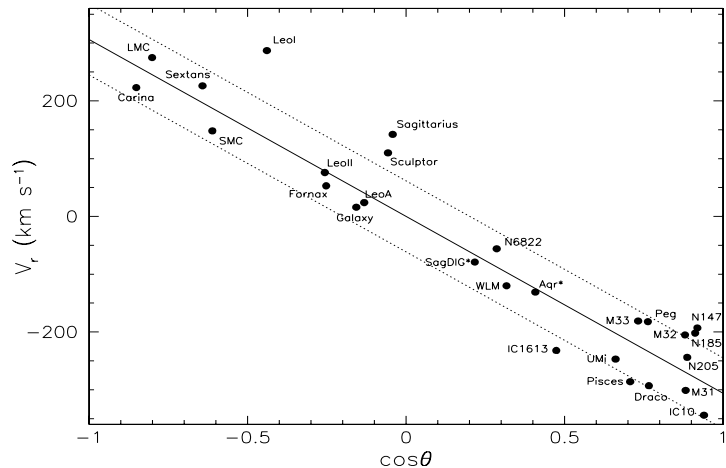

**Figure 4:** Radial velocity versus cosine of the apex distance for LG members. The plotted regression line corresponds to a solar velocity of 306 km s$^{-1}$ at cos θ = -1.0. The dotted lines lie 61 km s$^{-1}$ above and below the adopted regression line. Note the large deviations of the Sagittarius dwarf (which is interacting with the Galaxy), and of the high-velocity Leo I system.

Substituting $M_{LG} = (2.3 \pm 0.6)$ x $10^{12} \, M_\odot$ and $T_o = 14 \pm 2$ Gyr yields $R_o = 1.18 \pm 0.15$ Mpc.  The small uncertainty in $R_o$ is, of course, due to the fact that the error in $R_o$ depends on the cube root of the adopted LG mass.  Because the mass distribution in the LG is clumpy the actual shape of the zero-velocity surface of the LG will deviate from a sphere. From the data in Table 1 the total luminosity of the Local Group is $M_V$ = -22.0; 86% of which is provided by M 31 and the Galaxy. The main contributors to the uncertainty of $M_V$ are (1) the large error bars for the integrated luminosity of the Milky Way system, and (2) the fact that no (uncertain!) internal absorption corrections were applied to the observed luminosity of the almost edge-on Andromeda galaxy. From the total mass and integrated luminosity of the LG one obtains a mass-to-light ratio $M/L_V$ = 44 ± 12 (in solar units). Including internal absorption corrections for M 31 would decrease the value of $M/L_V$ .

Using the observed radial velocities of the members of the Andromeda subgroup of the LG Courteau & van den Bergh (1999) derived a total subgroup mass of (13.3 ± 1.8) x $10^{11}$ $M_\odot$ ; the lower and upper bounds corresponding to the virial and projected mass (Heisler, Tremaine & Bahcall 1985) estimates, respectively. This mass is slightly larger than the value (8.6 ± 4.0) x $10^{11}$ $M_\odot$ , that Zaritsky (1999) finds for the Milky Way subgroup. The combined mass of the Galactic and M 31 subgroups of the LG is therefore (21.9 ± 4.4) x $10^{11}$ $M_\odot$ . This value is, within its errors, identical to the (23 ± 6) x $10^{11}$ $M_\odot$ mass for the entire Local Group. This result suggests that most of the mass of the LG is tied up in the M 31 and Galaxy subgroups.  This conclusion would, however, have to be modified if Leo I is not a physical member of the Galactic subgroup of the LG.

## 6.    The luminosity distribution of Local Group galaxies

Figure 5 shows a plot of the luminosity distribution for the Local Group

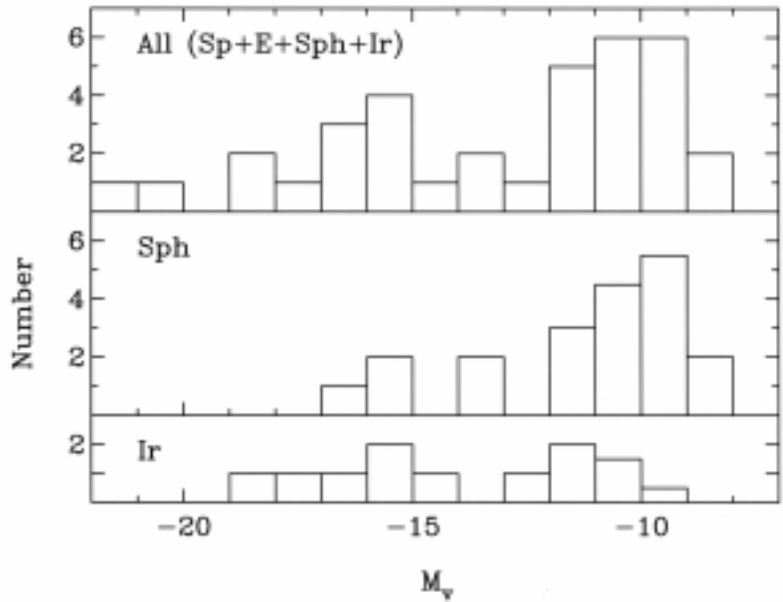

Figure 5        Luminosity distribution for LG galaxies.  The figure shows that the LG irregular galaxies have a shallower luminosity function than do the LG

dSph galaxies.  The drop below $M_V$ = -10 is due to incompleteness of the data.

members that are listed in Table 1.  The figure shows that the luminosity function of LG galaxies rises slowly towards fainter magnitudes.  The decrease below $M_V$ = -10 is, no doubt, due to incompleteness of the data.  This incompleteness is illustrated in Table 3, which shows that fewer dim galaxies fainter than $M_V$ = -10

Place Table 3 here

are known in the distant Andromeda subgroup than in the nearer Galactic subgroup. Inspection of Figure 5 shows that the slope of the luminosity distribution for dIr galaxies is much flatter than that for dSph galaxies. The luminosity distribution of the LG dSph galaxies is therefore more similar to that observed in rich clusters, than is to that for dIr galaxies. This result is, of course, not unexpected because early-type galaxies constitute the dominant population in rich clusters.

Evidence for a deficiency of galaxies with low surface brightnesses is also provided by Figure 6 [based on data in Mateo (1998)], which shows a deficiency of low surface brightness dwarf spheroidal companions of the Galaxy at distances $R_{GC} > 100$ kpc. The absence of distant low surface brightness Galactic satellites is probably due to the fact that such objects cannot be discovered as concentrations of very faint individual stars on Schmidt sky survey images.

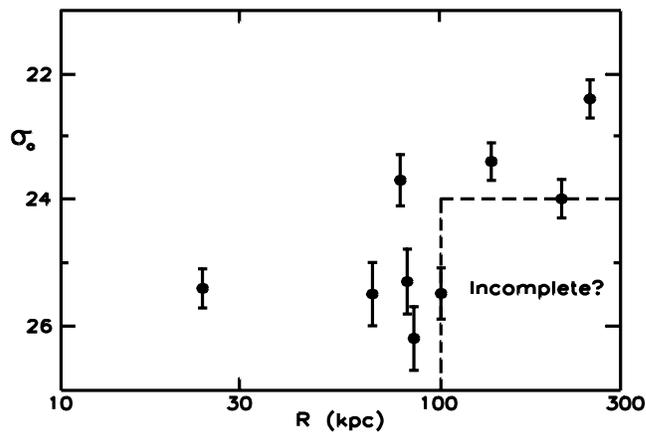

**Figure 6:**     Central brightness distribution of dSph companions to the Milky Way as a function of Galactocentric distance. The apparent correlation between $\sigma_0$ and $R_{GC}$ is probably due to the fact that it is difficult to discover unresolved very low surface brightness dSph galaxies beyond $R_{GC} = 100$ kpc. $\sigma_0$ is measured in mag arcsec$^{-2}$ .

The data on the luminosity distribution of LG galaxies may be fit to a luminosity distribution of the form (Schechter 1976)

$$\varphi(L) = \varphi^* \exp (-L/L^*) \times (L/L^*)^{\alpha} \, , \qquad (3)$$

in which $L^*$ is a characteristic luminosity at the transition between a power-law at faint magnitudes and an exponential cut-off at bright magnitudes. The number of luminous LG members is too small to determine $L^*$ with confidence. From maximum likelihood fits Pritchet & van den Bergh (1999) find that the LG data may be fit with a coefficient $\alpha = -1.1 \pm 0.1$, in which the exponent is not found to be very sensitive to the exact value down to which the luminosity distribution is assumed to be complete. The probability that the faint end of the LG luminosity distribution has a slope steeper than $\alpha = -1.3$ is less than one percent. This shows that the luminosity function of the Local Group differs significantly from those of rich clusters (Trentham 1998a,b), which have steep luminosity functions.

Figure 7 shows a comparison between the cumulative (integrated)

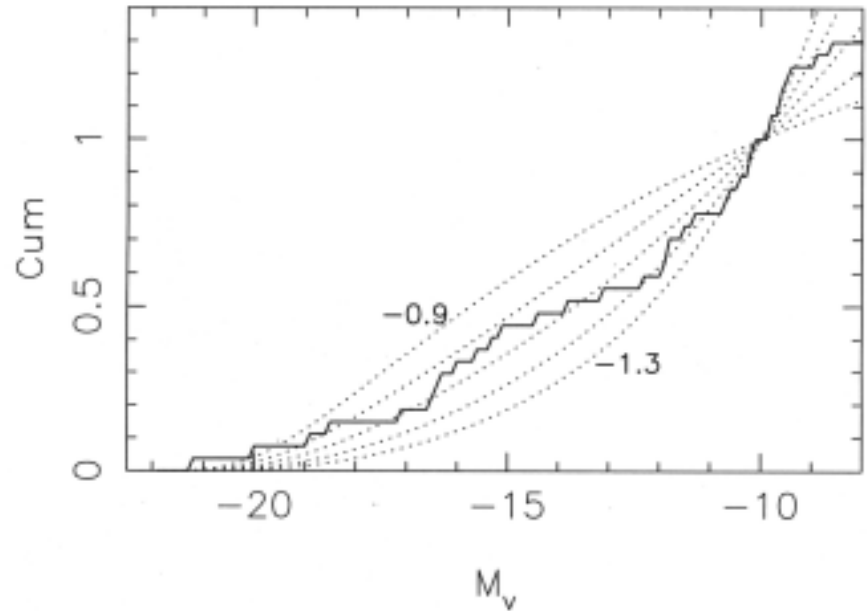

Figure 7    Comparison of the integrated luminosity distribution of the Local Group with Schechter functions having faint end slopes in the range  -1.3 ≤ α ≤ -

0.9  (Pritchet & van den Bergh 1999).  The figure shows that the slope of the LG luminosity distribution steepens towards fainter luminosities.  In the figure N(<$M_V$ )/N(<-10) is denoted by "Cum".

luminosity function of the LG and Schechter functions with various slopes normalized at $M_V$ = -10.  The figure shows that the LG luminosity distribution steepens towards faint luminosities.  The faint end of the LG luminosity distribution is seen to fit a Schechter function with α = -1.3.

# 7.    Properties of Local Group galaxies

Figure 8 shows a plot of luminosity versus metallicity for Local Group

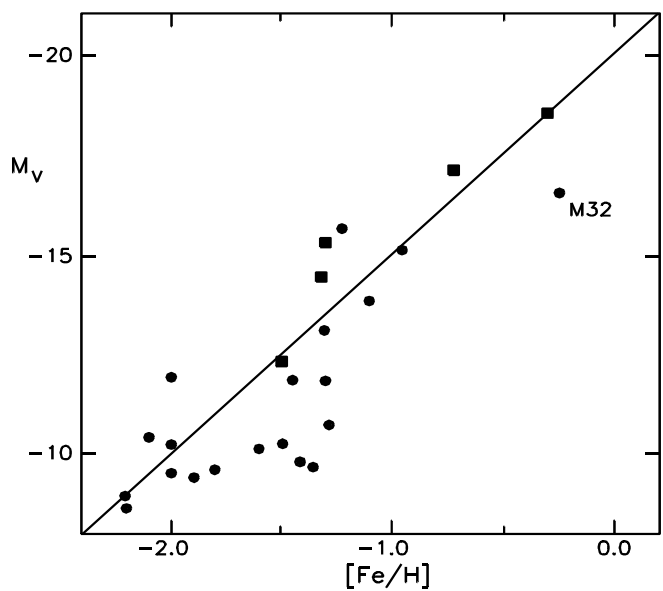

**Figure 8:**    Relation between metallicity [Fe/H] and luminosity $M_V$ for Local Group galaxies.  Note the close correlation between metallicity and luminosity.
The reason why M 32 deviates from this relation is discussed in the text.  The line plotted in the figure is given by Equation (4).

galaxies.  The data for individual LG members were taken from van den Bergh

(2000).  The figure shows metallicity growing by a factor of  ~ 100 as the

luminosity increases by a factor of 10 000.  Most of this luminosity dependence of

metallicity is probably due to the fact that low-mass galaxies are not able to retain

metals formed in supernovae, and ejected by supernova remnants and stellar

winds (Dekel & Silk 1986).  The fact that very few globular clusters show

evidence for internal enrichment indicates that almost all metals are lost from

systems with masses less than a few x $10^5$ $M_\odot$ .  It is of interest to note that M 32

appears to be subluminous for its metallicity.  It was first suggested by Faber (1973) that this effect is probably due to the fact that a significant fraction of the mass of M 32 was tidally stripped by interactions with M 31.  It is also noted that Ir galaxies (which are plotted as filled squares in Figure 8) lie systematically above galaxies of later types (which are shown as filled circles).  Presumably, the dIr galaxies will move down to the region now occupied by dSph galaxies after their young massive stars have evolved and faded.  The line drawn in Figure 8 is the relation

$$M_V = 20 - 5[Fe/H]. \tag{4}$$

Data compiled by Richer, McCall & Stasiński (1998) show a broadly similar increase of $\log(O/H)$ as galaxies become more luminous.

Figure 9 shows a plot of $M_V$ versus disk scale-length for Local Group

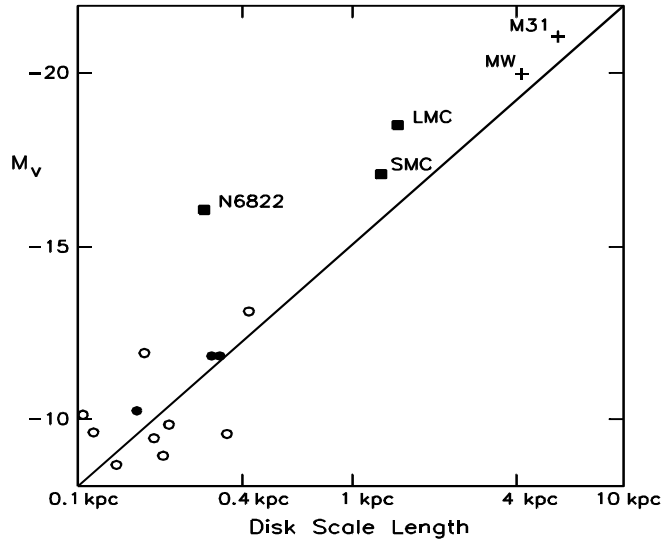

galaxies, that is based on the compilation in van den Bergh (2000).  Not

unexpectedly, the most luminous galaxies tend to have the largest disk scale-

lengths.  Probably due to the effects of luminosity evolution the irregular galaxies

in this figure lie above the trend line for the majority of data points.  It is also of

interest to note that the dwarf spheroidals located in the Galactic subgroup of the

Local Group, (open circles) follow the same trend as do the those in the

Andromeda subgroup (filled circles).

## 8.    Interstellar material in the Local Group

The idea that the LG might contain significant amounts of intra-cluster gas was introduced by Kahn & Woltjer (1959).  These authors invoked the presence of vast amounts of (difficult to observe!) hydrogen gas at $5 \times 10^5$ K to account for the "missing mass" that seemed to be indicated by dynamical arguments.  More recently (Oort 1966, 1970) discussed the possibility that capture of pristine neutral hydrogen gas in high-velocity intra-cluster clouds might increase the mass of the galaxy by ~ 1% per Gyr.  Probable examples of such inflows are discussed by Wakker & van Woerden (1997).  On the other hand Morris & van den Bergh (1994) have argued that some interstellar gas will be returned to intra-galactic space by tidal interactions between galaxies.  The importance of such tidal interactions is demonstrated by the M 81 group, in which $1.4 \times 10^9$ $M_\odot$ of neutral hydrogen (Appleton, Davies & Stephenson 1981) is found to be located outside the Holmberg radii of NGC 3031 (M 81), NGC 3034 (M 82) and NGC 3077. Within the Local Group the Magellanic Stream (Mathewson & Ford 1984), and the more recently formed bridge between the LMC and the SMC, provide the most spectacular evidence for tidal gas loss from galaxies.  Furthermore Heiles (1987) has argued that supernova-induced fountains will eject some galactic gas back into intergalactic space.

Bland-Hawthorn & Maloney (1998) have suggested that high-velocity clouds might be observed in Hα , if their surface layers are ionized by ultraviolet radiation. Wakker, van Woerden & Gibson (1999) point out that observations of the strength of Hα in the compact high-velocity clouds discussed by Blitz et al. (1998) could establish if they are truly intra-galactic, or if they are associated with individual star forming LG members. In the latter case one would expect their Hα radiation to be strongly enhanced by UV photons escaping from regions of star formation.

Continuing infall of clumps of gas is an inevitable consequence of hierarchical merger models for structure in the Universe. Blitz et al. regard compact high-velocity clouds as examples of such infalling gas that is bound by "mini-halos" of cold dark matter. According to Blitz et al. such objects typically have hydrogen masses of $\sim 3 \times 10^7$ $M_\odot$ , total masses of $\sim 3 \times 10^8$ $M_\odot$ , and are located at distances of $\sim 1$ Mpc. However, it is not yet clear if such (presently still hypothetical) objects will be surrounded by enough ionized gas to shield their neutral hydrogen from the ionizing background. Figure 10 shows a plot of the velocity.  The data suggest that many of the compact HV clouds may still be falling into the Local Group.

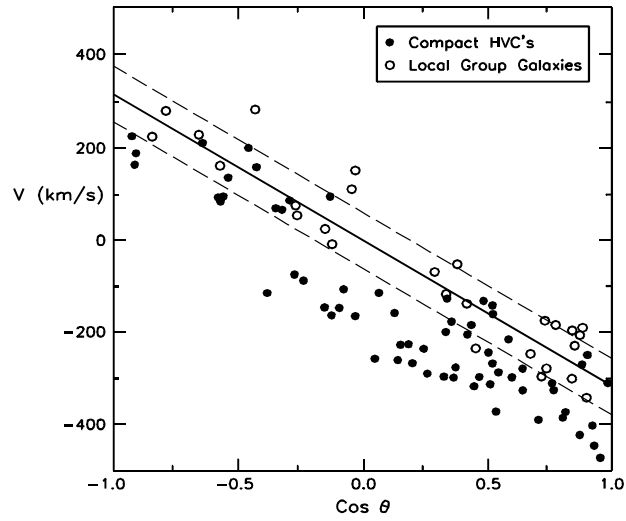

**Figure 10:**    Plot of Local Group galaxies (open circles) and compact HV clouds (filled circles) in a V versus cos θ plot, in which V is the observed radial

locations of the compact H I clouds of Burton & Braun (1999) in a $V_r$ versus apex angle plot.  Also shown for comparison are the positions of LG members in this same plot.  The figure shows that the compact H I clouds are typically blueshifted by ~ 100 km s$^{-1}$ relative to the LG galaxies.  Taken at face value this result suggests that the compact H I clouds are still falling into the Local Group.

## 9.    Group properties

A summary of the properties of the Local Group is given in Table 4.  The

Place Table 4 here

data in this table are drawn from the previous sections, from van den Bergh (2000), and from the references cited in the table. Information on individual LG members will be given in Sections 10 - 19.

## 10. The Andromeda galaxy

### 10.1 Introduction

The Andromeda galaxy (= M 31 = NGC 224) is the most luminous object in the Local Group. The fact that it is so bright immediately makes one suspect that it might have formed by mergers of lesser galaxies (Freeman 1999). This suspicion is strongly supported by the star counts of Pritchet & van den Bergh (1994) which, in conjunction with previous work, show (see Figure 11) that the

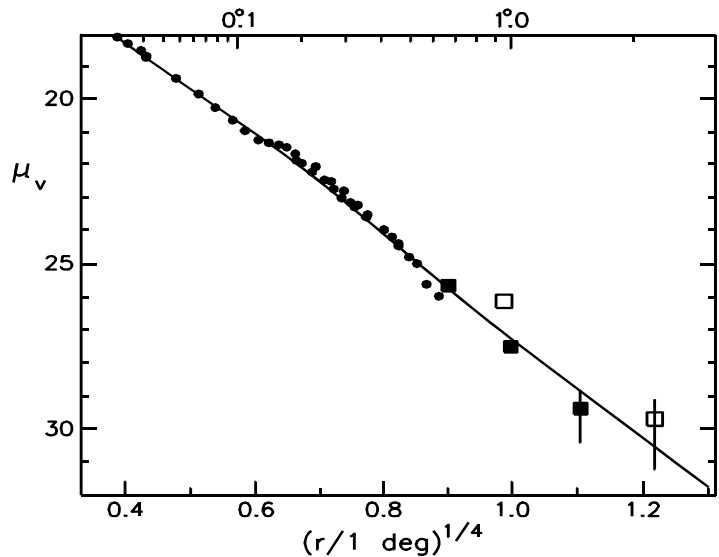

**Figure 11:**     The radial surface brightness profile of M 31. Filled squares are from the observations of Pritchet & van den Bergh (1994) along the minor axis of the Andromeda galaxy. The two open squares represent observations of fields along the major axis (which have been contaminated by disk stars).

Dots are observations by other authors.

radial profile of the spheroidal component of the Andromeda galaxy can be represented by a de Vaucouleurs $R^{\frac{1}{4}}$ law over an enormous range in R.  Such a profile is predicted to result from tumultuous mergers between galaxies of comparable mass, that produce violent relaxation   (White 1979, Barnes 1990).  It is of interest to note that the halo density in M 31  (scaled for galaxy size) is an order of magnitude higher than it is in the Galaxy.  This suggests that the Galaxy might mainly have formed from the collapse of a single protogalaxy, with later infall (and capture) of <u>minor</u> bits and pieces of the type described by Searle & Zinn (1978).  The notion that M 31 and the Galaxy had a very different initial evolutionary history is supported by the observation that, compared to its abundance in Galactic globulars, nitrogen appears to be enhanced in M 31 globular clusters (Burstein et al. 1984, Ponder et al. 1998, Worthy 1998).

The second striking feature of M 31 is that both its globular clusters (van den Bergh 1969), and its halo (Mould & Kristian 1986, Pritchet & van den Bergh 1988), are unexpectedly metal-rich.  This strongly suggests that most of the star formation, and production of heavy elements in the Andromeda Galaxy, occurred long ago.  The hypothesis that M 31 had an active past, but is now relatively quiescent, is supported by observations of the integrated luminosities and colors of its star clusters (see Figure 12).  This figure shows that the Andromeda galaxy

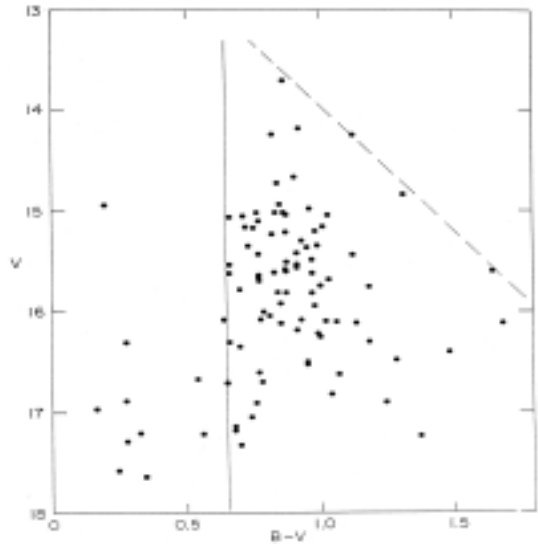

Figure 12    Color-magnitude diagram for star clusters in M 31. Open clusters have B-V < 0.66. The dashed line is a reddening line with $A_V/E(B-V) = 2.5$.

The figure shows that M 31 contains far more old red clusters than young blue ones.

contains many old red clusters, and relatively few young blue ones.

## 10.2    The nucleus of M 31

Embedded within the nuclear bulge of M 31 is a bright, metal-rich, and dynamically distinct nucleus. This nucleus (= BD + 40° 148) has V = 12.6 ± 0.3, corresponding to $M_V$ = -12.0. This luminosity is so high that no less than ~ 60 typical globulars would have to have been dragged into the nucleus by dynamical friction (Tremaine, Ostriker & Spitzer 1975) to account for its luminosity. Another argument against the hypothesis that the nucleus of M 31 was entirely built up by the merger of globulars is that its metallicity is higher than that of all but a very few of the globular clusters associated with the Andromeda galaxy (van den Bergh 1969). Lauer et al. (1993) have used the <u>Hubble</u> <u>Space</u> <u>Telescope</u> to show that the nucleus of M 31 is double, with a separation of $0\overset{''}{.}49$ (1.8 pc),in which the fainter component ($P_2$) is situated at the geometric center of the bulge. $P_2$ has a half-light radius of ~ 0.2 pc and contains a black hole of mass ~ 5 x $10^7$ $M_\odot$ (Dressler & Richstone 1988, Kormendy 1988). Tremaine (1995) has made the attractive suggestion that the brighter component $P_1$ of the nucleus is produced by stars, in a small nuclear disk, that linger near apoapsis. The observation (King, Stanford & Crane 1995) that $P_2$, which is situated at the center of the M 31 bulge isophotes, has a larger UV upturn than $P_1$ might be explained by assuming that the nucleus did tidally capture a <u>few</u> globulars with blue horizontal branches. The discovery (Lauer et al. 1996) that the Virgo galaxy NGC 4486B also has a binary nucleus shows that such objects are not excessively rare. Crane, Dickel & Cowan (1992) have found that the nucleus of M 31 contains a (possibly variable) radio source with a luminosity ~ 1/30 of that of Sgr A* in the Galactic nucleus. This low level of activity suggests that the black hole in the nucleus of M 31 is presently fuel-starved.

10.3    The bulge of M 31

The nuclear bulge contributes ~ 30% of the visual light of the Andromeda galaxy. Observations of the integrated light of the bulge of M 31 (Morgan 1959, McClure & van den Bergh 1968) show that metal-rich stars make a major contribution to its luminosity (see Figure 13). This disproves the notion (Baade 1944a,b) that Population II stars are the dominant population in the bulge of M 31.  The high N/H and O/H ratios for the gas in the bulge of M 31 (Rubin, Kumar & Ford 1972) show that this gas must have been ejected by evolving stars and supernovae.

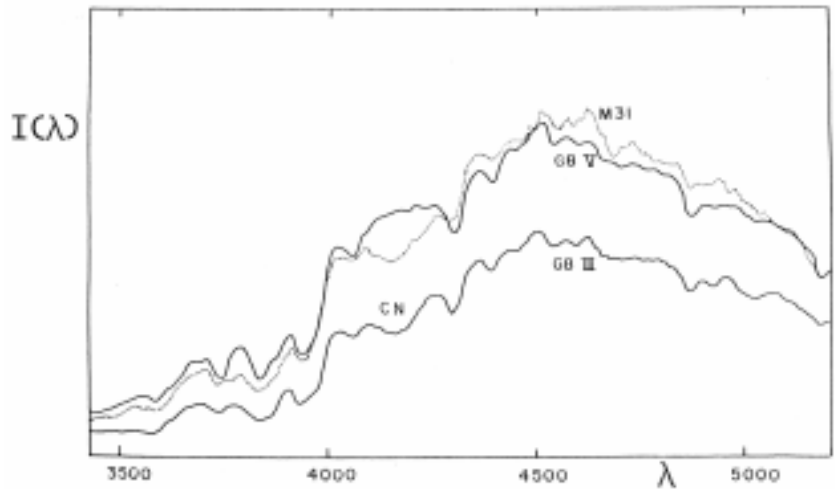

Figure 13      Spectral energy distribution of the central region of the M 31 bulge (van den Bergh & Henry 1962), compared to those of G8V and G8III stars of normal metallicity.   The figure shows that the light emitted by the central bulge of M 31 is  dominated by metal-rich giant stars.

10.4      The disk of M 31

Tinsley & Spinrad (1971) were able to show that the light emitted by the disk of M 31 is dominated by an old stellar population.  The fact that the scale-length of this exponential disk decreases from 34´ (7.5 kpc) in U, to 20´ (4.4 kpc) in the K-band, shows that the younger disk population grows in importance with increasing radius.  The first hint of such an effect was noted by Babcock (1939), who found that the Balmer lines in the disk spectrum strengthen with increasing distance from the nucleus.  Spectroscopic observations of H II regions and supernova remnants  (Blair, Kirshner & Chevalier 1981) show that the disk of M 31 exhibits a significant radial metallicity gradient.

Digital stacking of Palomar Sky Survey plates by Innanen et al. (1982) demonstrates that the outermost part of the disk of the Andromeda galaxy is warped.  Radio observations show that the neutral hydrogen disk of M 31 is warped in the same direction as the optical disk.  However, the 21-cm isophotes appear to remain close to the fundamental plane of M 31 out to somewhat greater distances from the nucleus than is the case for the optical isophotes.

Embedded within the disk of M 31 is a two-armed spiral structure.  This structure is most clearly outlined by dust lanes in the inner part of M 31, and by associations of OB stars in its outer regions.  Van den Bergh (1964) has mapped 188 OB associations in M 31.  The associations in Andromeda appear to be larger than those in the Galaxy, M 33 and the Magellanic Clouds.  The reason for this apparent difference is that the low density of young stars in the disk of M 31 allows one to trace the associations out to greater distances, than is possible in the rich star fields in the Milky Way, M 33 and the Clouds of Magellan.  The spiral structure, as outlined by the OB associations and H II regions, reaches its maximum strength in a broad ring with R ~ 10 kpc.  The integrated H-$\alpha$ flux of the M 31 disk shows (Walterbos & Braun 1994) that the current rate of star formation is only $\sim 0.35\ M_\odot$ per year. Dame et al. (1993) find that the molecular gas in the Andromeda galaxy is concentrated in the 10 kpc ring that exhibits the maximum display of OB stars and H II regions.

### 10.5   The halo of M 31

The halo of M 31 contains $\sim 400$ globular clusters (Fusi Pecci et al. 1993), field RR Lyrae variables (Pritchet & van den Bergh 1987), and old red giant stars (Mould & Kristian 1986). The fact that the Andromeda galaxy contains 2-3 times as many globulars as the Milky Way system is probably due to both the higher luminosity of M 31, and to the fact that it is of earlier morphological type than the Galaxy. Hubble Space Telescope observations of horizontal branch stars in M 31 globular clusters of differing metallicity (which are all located at the same distance) by Fusi Pecci et al. (1996), appear to favor the view that the luminosities of RR Lyrae stars are not strongly dependent on their metallicities. It would, however, be important to obtain observations of additional globular clusters in the halo of M 31 to strengthen the determination of the relationship between $M_V$ (RR) and [Fe/H], which is of critical importance to the extragalactic distance scale problem.

The radial density distribution of stars in the outer halo of M 31 may be modeled by a power law of the form $\rho(R) \propto R^{-5}$, whereas the stellar density distribution in the outer halo can be represented by the power law $\rho(R) \propto R^{-3}$ (Pritchet & van den Bergh 1994). This shows that the globular cluster population of M 31 is more extended than its stellar population. The projected stellar distribution may be approximated by a de Vaucouleurs $R^{1/4}$ law over the range R = 200 pc to R ~ 20 kpc. Using two-color observations of individual giant stars it should be possible to extend the density profile of the M 31 halo to considerably larger distances. Van den Bergh & Pritchet (1992) find that the mean metallicity for stars in the inner halo of M 31 has the surprisingly high value [Fe/H] ~ -1.0. Furthermore no abundance gradient is found for halo stars over the range 9 < R (kpc) $\lesssim$ 20. Beyond R ~ 5 kpc globular clusters also appear to exhibit no abundance gradient. Possibly violent relaxation, following the merger of two massive ancestral galaxies, accounts for the presence of relatively metal-rich stars and clusters in the halo of M 31.

## 11.    The Galaxy

### 11.1    Introduction

Proof of the suspicion that the Milky Way might be an isolated stellar system, rather than the entire Universe, was first provided by Shapley (1918a,b), who used the distribution of globular clusters to show that the center of our galaxy was associated with the rich star clouds near the boundaries of Scorpio and Ophiuchus. This idea was subsequently confirmed by Lindblad (1927) and Oort (1927, 1928), who showed that (a) the Milky Way system is in differential rotation around a center, which coincides with that of Shapley's globular cluster system, and (b) that the system of globular clusters and the "high-velocity" stars rotate more slowly than does the Galactic disk. The notion that the Milky Way system was a spiral galaxy received its ultimate confirmation when Morgan, Whitford & Code (1953) were able to identify the nearest spiral arms, and when Westerhout (1954) and Schmidt (1954) used 21-cm line observations, which penetrated the huge dust clouds in the plane of the Milky Way, to map the global spiral structure of our galaxy. A number of lines of evidence now suggest that the Galaxy is of Hubble stage bc. The discovery of a short bar in the central region of the Milky Way (Blitz & Spergel 1991) shows that a morphological type S(B)bc is probably appropriate for the Galaxy.

11.2    The Galactic nucleus

The Galactic nucleus, which is obscured by $\sim 30$ magnitudes of absorption at visual wavelengths, coincides with the radio source Sgr A*, that is located at $\ell = 17^h 45^m 39\overset{s}{.}9$, b = $-29° 00' 28''$ (J2000). Backer & Sramek (1982) find that its proper motion is consistent with the hypothesis that Sgr A* is at rest in the Galactic nucleus. Available evidence (Reid 1993) suggests that the distance $R_o$ to the Galactic nucleus is 8.5 ± 0.5 kpc.  From K-band observations

with the Keck 10-m telescope Ghez et al. (1998) find that both the stellar surface

brightness, and the highest observed stellar velocity dispersion, are consistent

with the presence of a massive black hole.  From proper motion observations of

individual stars Eckert & Genzel (1997) derive a mass M = (2.45 ± 0.14) x $10^6$

$M_\odot$ for the region within 0.015 pc of the nucleus.   Particularly strong evidence for

the existence of such a black hole is provided by the observation that a star at 0".19

(0.007 pc)  from Sgr A* has a space motion ≥ 1500 km s$^{-1}$.

The nucleus of the Galaxy is surrounded by a star forming disk with a

radius of 115 pc.  Embedded within this disk is the remarkable "arches" cluster G

0.121 + 0.017, which contains over 100 massive young O-type stars (Serabyn,

Shupe & Figer 1998).  The radius of this cluster is only ⅓ that of R 136, at the

center of the 30 Doradus complex, and its stellar density is higher than that in R

136.  These data suggest that conditions within a few tens of pc of Sgr A* are

particularly favorable for the formation of massive dense clusters.  The recent

detection of γ-rays from within  ~ 1° of the Galactic center (Melia, Yusef-Zadeh

& Fatuzzo 1998) shows that the emission of this region extends to very high energies.

### 11.3    The nuclear bulge

Since Baade (1944a,b) believed the nuclear bulges of galaxies to be composed of globular cluster-like stars of Population II, he (Baade 1951) looked for (and found!) RR Lyrae variables in the low-absorption window at $\ell = 1.0$ and $b = -3°.9$.  However, Morgan (1959) showed that the integrated spectrum of the star cloud seen in "Baade's Window" was dominated by metal-rich giants.  This demonstrated that metal-poor stars of Population II were only a minor constituent of the bulge population.  From an analysis of the spectra of K-type giant stars in Baade's window Sadler, Rich & Terndrup (1996) found that the bulge stars sampled in Baade's window have $<[Fe/H]> = -0.11 \pm 0.04$.  There are two reasons why the true mean metallicity is probably higher than this value.  In the first place the line of sight at $b = -3°.9$ intersects the bulge at $Z = -580$ pc, where the mean metallicity is expected to be lower than it is at $Z = 0$ pc.  Secondly the analysis by Sadler et al. was restricted to K-type giants, whereas the most metal-rich stars are expected to spend a significant fraction of their lifetimes as cool M-type giants (Spinrad, Taylor & van den Bergh 1969).  Idarte et al. (1996) find $[Mg/Fe] = +0.45$ for the giants in Baade's Window.  Taken at face value this result suggests that metal enrichment in the bulge was dominated by fast-evolving supernovae of Type II, rather than by iron-rich gas ejected from more slowly evolving progenitors of supernovae of Type Ia.  However, McWilliam & Rich (1994) find $<[CN/Fe]>$ to be closer to solar in bulge stars.  If this conclusion is correct then these objects must differ, in some respects, from the stars that dominate the light of elliptical galaxies. A similar problem was noted in Section 10, where it was found that the N abundance in M 31 and its globular clusters, is higher than it is in the Galaxy. Another unexplained observation (McWilliam & Rich 1994) is that [Ca/Fe] and [Si/Fe] in bulge stars appear to be close to values observed in Galactic disk stars. For a detailed discussion of these problems the reader is referred to Worthy (1998). From radial velocities of typical metal-rich stars in the Galactic bulge Minniti (1996) finds a rotational velocity V(rot) = +66 $\pm$ 5 km s$^{-1}$. However, he finds that metal-poor stars in this field with [Fe/H] < -1.5 have V(rot) = -6 $\pm$ 20 km s$^{-1}$, i.e. they do not participate in Galactic rotation. From a rather noisy color-magnitude diagram, that extended down to the main sequence turnoff, Terndrup (1988) concluded that the bulk of the stars in Baade's window have ages in the range 11-14 Gyr.

It is still not clear how the Galactic bulge formed. Carney, Latham & Laird (1990) have suggested that it was made from low angular momentum gas, that was left over after most star formation had ended in the halo. Alternatively Combes et al. (1990) believe that bulges form from bars. However, it is not yet clear if this suggestion is consistent with the observation of both radial and vertical abundance gradients in galactic bulges. The existence of a small nuclear bar (Kuijken 1996), that is embedded within the Galactic bulge, might have interesting dynamical consequences. Lerner, Sundin & Thomasson (1999) have pointed out that stars initially situated just outside such a bar will be strongly perturbed on each orbit; each pericenter passage boosting the star into a more elongated orbit. From numerical experiments they find that this process is fast, and can boost stars out to distances corresponding to 10 bar lengths in less than 2 Gyr. Perhaps such a process can account for the fact that the oldest known open cluster NGC 6791, which has an age $T = 8.0 \pm 0.5$ Gyr (Chaboyer, Green & Liebert 1999), has such a high metallicity ([Fe/H = +0.4 ± 0.1). The fact that the motion of this cluster (Scott et al. 1995) lags circular disk motion by more than 60 km s$^{-1}$ appears consistent with the idea that this metal-rich cluster was ejected outwards from the bulge into the solar neigborhood.

### 11.4    The Galactic disk

The evolutionary relationships between the Thin Disk, the Thick Disk and the Galactic halo remain a subject of lively controversy. In particular it is not yet clear if the evolution from protogalaxy to disk was continuous, or if there was an extended hiatus between the halo and disk phases of Galactic evolution (Berman & Suchkov 1991). Furthermore it is not known if the Thick Disk and Thin Disk represent distinct evolutionary phases, or if they simply represent the extremes of a continuous evolution. A number of lines of evidence suggest that the oldest stars in the Thick Disk have ages $\geq 10$ Gyr. No old Thin Disk clusters are found to have $R_{GC} < 7$ kpc (Friel 1995). The reason for this may be that interactions with giant molecular clouds, which are numerous inside the solar radius, destroy old open clusters in a few Gyr or less (van den Bergh & McClure 1980). The fact that old open clusters have not been destroyed in the outer Galactic disk appears to militate against the suggestion (Pfenniger, Combes & Martinet 1994, Combes & Pfenniger 1997) that, so far undetected, massive cold H I clouds are present in the outer disk. Observations of both H II regions, and of B stars, show a smooth decline in the oxygen abundance over the range $5 \lesssim R_{GC} \lesssim 15$ kpc. The most distant known open cluster Berkeley 21 (Tosi et al. 1998) at $R_{GC} = 14.5$ kpc, which has [Fe/H] ~ -1.0, is at least as metal-poor as the Small Magellanic Cloud. Edvardsson et al. (1993) have found that the metallicity of stars with $7 \lesssim R_{GC} \lesssim 13$ kpc, that have ages $\lesssim 10$ Gyr, exhibit no systematic dependence of metallicity on age. The most straightforward interpretation of this observation is that infall of metal-poor pristine gas has, almost exactly, compensated for the increase of metallicity resulting from the effects of nucleosynthesis and ejection of processed material back into the interstellar medium by supernovae, novae and stellar winds.

11.5    The Galactic halo

Toomre (1977) argued that major galaxies formed by "merging of sizable bits and pieces (including quite a few lesser galaxies)".  Searle & Zinn (1978) elaborated on this theme by suggesting that our own Milky Way system had been assembled in this way, rather than by the monolithic collapse envisioned by Eggen, Lynden-Bell and Sandage (1962).  After a great deal of discussion a consensus seems to be emerging (van den Bergh 1996c) in which monolithic collapse is regarded as the principal mode of formation inside the solar circle; whereas capture of bits and pieces was the dominant factor in forming the Galactic halo at $R_{GC} \geq 8$ kpc.  As a result of such infall the outer halo is, in a dynamical sense, presently not well mixed (Majewski, Munn & Hawley 1996).  This conclusion has, more recently, been supported by HIPPARCOS proper motion and parallax observations (Chen 1998).  Prima facie evidence for mergers is provided by the Sagittarius dwarf (Ibata et al. 1994), which is presently being disrupted by the Milky Way.  However, the low frequency of "young" globular clusters (which appear to form in dwarf spheroidals) in the Galactic halo places severe restrictions on the fraction of the total halo luminosity that could have been derived from mergers with dwarf spheroidal galaxies.  According to Richer et al. (1996) the following nine globulars are known to be "young", i.e. at least  ~ 3 Gyr younger than the majority of globulars:  IC 4499, Eridanus, Arp 2, Pal 1, Pal 3, Pal 4, Pal 12, Ruprecht 106, and Terzan 7.  Of these nine objects two (Arp 2 and Ter 7 appear to be associated with the Sagittarius dwarf.  Furthermore Irwin (1999) has suggested that Pal 12 may have been tidally detached from Sagittarius. This indicates that ~ 1/3 of all young globulars may be associated with Sagittarius. If so, one would not expect mergers to have contributed stars with a total luminosity of more than about three times the luminosity of Sagittarius ($M_V$ ~ -14) to the halo, for which $M_V$ = -18.4 (Suntzeff 1992). Preston, Beers & Shectman (1994) estimate that young accreted material might account for 4-10% of the Galactic halo.

Traditionally (Kinman 1959, Zinn 1985) the Galactic globular cluster system is regarded as consisting of metal-rich disk, and metal-poor halo components. However, Woltjer (1975), and inspection of Figure 14 suggests that the population of metal-rich clusters with [Fe/H] > -1.0 might more properly be regarded as being associated with the Galactic bulge, rather than with the disk.  A second striking feature of Figure 14 is that the bulge shows up clearly among the metal-poor globular clusters.

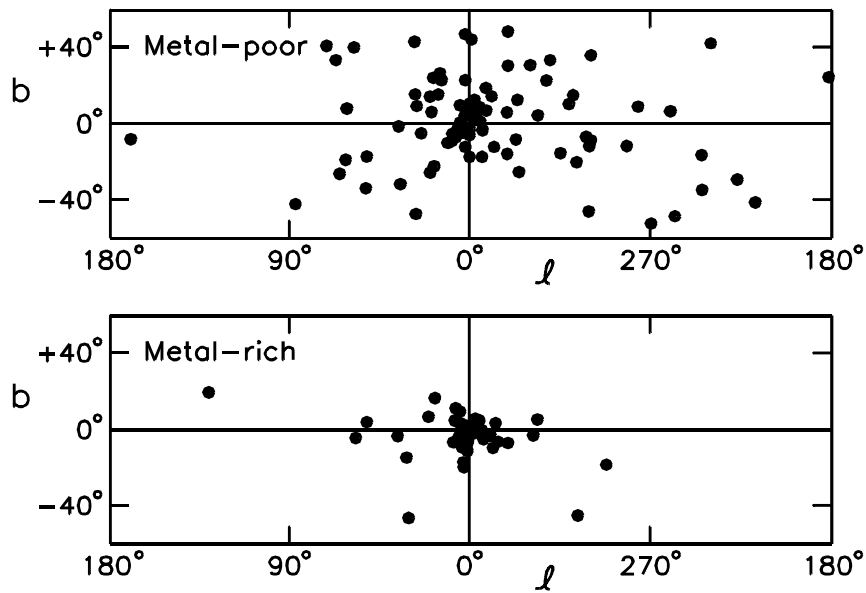

**Figure 14:** Comparison between the distribution of metal-poor ([Fe/H] < -1.0) and metal-rich ([Fe/H] > -1.0) globular clusters on the sky. Note the strong concentration of metal-rich clusters towards the Galactic center. Contrary to expectation the metal-rich clusters do not seem to outline the Thick disk.

Van den Bergh (1993) and Lee, Demarque & Zinn (1994) have pointed out that the Galactic globular clusters with -2.0 < [Fe/H] < -1.0 can be divided into two populations on the basis of their position in a plot of [Fe/H] versus horizontal branch population gradient (B-R)/B+V+R), in which B, V and R are the number of blue, variable, and red stars, respectively.  In a (B-R)/(B+V+R) versus [Fe/H] plot the globular clusters interior to the Sun appear to lie along a single curve (isochrone?).  Van den Bergh (1993) assigns these objects, which presumably formed during the collapse of the main body of the Galaxy, to his β population.  On the other hand globular clusters outside the solar circle, which constitute van den Bergh's α population, mostly lie below this curve in the (B-)/(B+V+R) plot.  This population may represent objects that were captured as they fell into the Galaxy at a later date.  Figure 15 shows a plot of $M_V$ versus $R_{GC}$ for Galactic globulars. The figure shows that all "young" clusters (shown as filled circles), which have ages that are at least 3 Gyr less than those of the majority of globulars, lie beyond R = 15 kpc. Furthermore the clusters in the outer halo are [with the exception of the very old metal-poor cluster NGC 2419 (Harris et al. 1997)]) seen to be systematically less luminous than are the globular clusters in the main body of the Galactic halo. Taken at face value these data suggest the possibility that the mean luminosity of globulars in the outer Galactic halo may, at their time of birth, have decreased with time, reaching values similar to those of typical open clusters after ~ 5 Gyr.

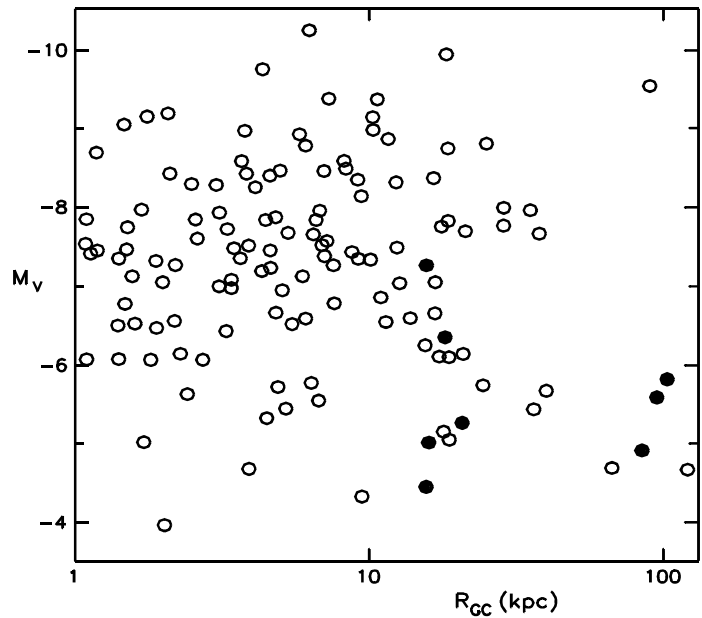

**Figure 15:** Plot of $M_V$ versus $R_{GC}$ for Galactic globular clusters. The figure shows that all "young" globular clusters (filled circles) are located at $R_{GC} > 15$ kpc, whereas "old" clusters (open circles) occur at all radii. It is of interest to note that globular clusters in the outer halo are, on average, fainter than those located in the main body of the halo.

One of the most curious features of the Galactic globular cluster system is that the half-light radii of clusters increase with Galactocentric distance (van den Bergh 1996c).  The fact that no small clusters exist far out in the halo, even though such objects would have easily survived, shows that this relation is not entirely due to tidal disruption of large clusters at small values of $R_{GC}$.  In other words large globular clusters tend to form far from the Galactic center, in what were, presumably, regions of rather low space density.  It is curious that the large halo cluster NGC 2419, situated at 0.1 Mpc from the Galactic center, has an age that is indistinguishable from that of the oldest clusters in the main body of the Galaxy.  In other words cluster formation seems to have started more-or-less simultaneously  ~ 15 Gyr ago in regions of very different density.

### 11.6    Comparison between the Galaxy and M 31

A comparison between the main characteristics of the Andromeda galaxy, and of the Milky Way system is shown in Table 5.  The data in this table show

Place Table 5 here

that the nuclear bulge of M 31 is significantly more luminous than that of the Galaxy; even though their effective radii appear to be similar.  Furthermore the velocity dispersion in the bulge, and the overall rotational velocity are larger in M 31, than they are in the Galaxy. Finally the present rate of star formation, and hence $L_{IR}$, is greater in the Milky Way system, than it is in the Andromeda galaxy.

## 12.    The Triangulum galaxy = M 33

M 33 (= NGC 598) is the third-brightest member of the Local Group. It has type Sc II-III and a luminosity $M_V$ = -18.9. The Triangulum spiral has four principal components: (1) an exponential disk, that contributes most of the luminosity of this galaxy, (2) a halo, that contains RR Lyrae stars and globular clusters, (3) a semi-stellar nucleus, and (4) perhaps a small nuclear bulge. The nucleus of M 33 has $M_V$ = -10.8, a half-light radius $r_h$ = 0.″35 (1.4 pc), and a velocity dispersion of only 21 ± 3 km s$^{-1}$ (Kormendy & McClure 1993). From these data the mass-to-light ratio $M/L_V$ ≲ 0.4 (in solar units). This shows that this nucleus does not contain a massive black hole. This conclusion is consistent with the notion that the masses of central black holes scale with those of the galactic bulges in which they are embedded. The nucleus of M 33 has a composite spectrum (van den Bergh 1976), with a late A-type being indicated by K/(H + Hε), F2-4 by λ 4226/Hγ, and F3-4 by CH/Hγ . From observations in the range λλ1200 - 3000 Gordon et al. (1999) conclude that this nucleus underwent a starburst in which $M_V$ ~ -15.5 at 10 Myr after outburst. The fact that M 33 has a well-developed halo, but that its bulge is so dim that its very existence is in doubt, shows that bulges are a distinct galactic component; not just the extensions of halos to small radii. The disk of M 33, which has a scale-length of ~ 2 kpc, is embedded within a much larger neutral hydrogen envelope. The sharp outer cut-off of this hydrogen disk (Kenney 1990) is believed to be due to ionization by the intergalactic UV radiation field. The hydrogen in the outer disk of M 33 is tilted by ~ 30° with respect to the inner disk. The origin of this tilt is not understood. Since M 33 is quite distant from M 31, it cannot be due to a recent tidal interaction between these two galaxies. Embedded within the disk of M 33 are numerous star clusters, OB associations and H II regions. The rate of star formation in this disk is ~ 0.04 $M_\odot$ per year. The scale-length for the surface density of associations (van den Bergh 1991), which is plotted in Figure 16, is 9.9 (2.4 kpc). This value is close to the ~ 2 kpc scale-length for the optical light in the disk of M 33.

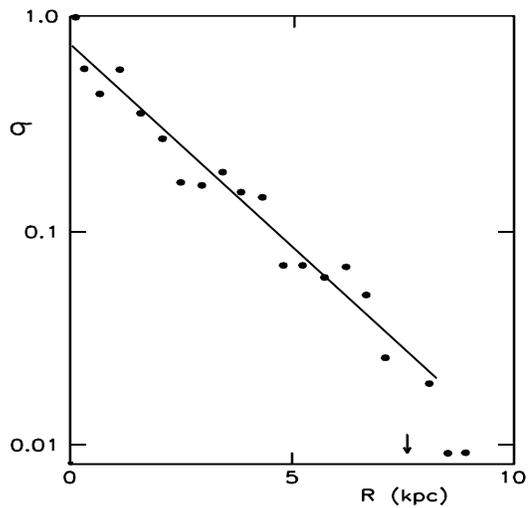

**Figure 16:** Normalized and rectified surface distribution of associations in M 33 (Ivanov 1978). For R < 8 kpc the observed density distribution has a scale-length of 9.9.

Figure 17 shows (Schommer et al. 1991) that the reddest M 33 clusters

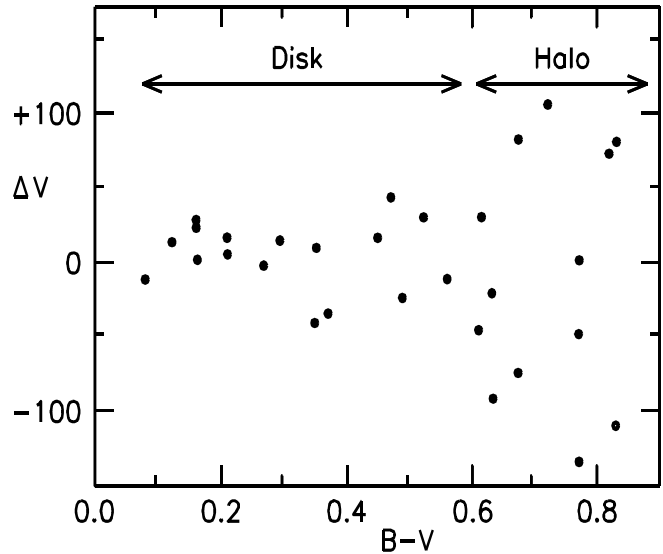

**Figure 17:**    Velocity difference $\Delta V = V(cluster) - V(disk)$ in km s$^{-1}$, versus integrated cluster color, from data by Schommer et al. (1991).  The plot shows that the old red clusters in M 33 have the kinematics expected for a halo population.

with B-V > 0.66 have halo kinematics with a velocity dispersion, relative to the

Population I rotation curve, of ~ 70 km s$^{-1}$.  On the other hand the rotation of the

bluer (younger) clusters with B-V < 0.66 is indistinguishable from that of the HI

gas and the H II emission regions.  Sarajedini et al. (1998) have recently used the

<u>Hubble</u> <u>Space</u> <u>Telescope</u> to obtain color-magnitude diagrams for 10 of the red

halo clusters in M 33.  These authors find that the M 33 halo clusters have

metallicities as low as [Fe/H] = -1.6.  However, the morphology of their color-

magnitude diagrams suggests that these clusters are a few Gyr younger than the

majority of globulars in the Galactic halo and in the Large Magellanic Cloud.  On

the basis of current ideas on galaxy evolution it is difficult to understand why the

<u>halo</u> clusters in M 33 appear to have formed more recently than the <u>disk</u> globular clusters in the LMC. For reviews on M 33 the reader is referred to Gordon (1969) and to van den Bergh (1991).

## 13.    The Large Magellanic Cloud

### 13.1    Introduction

To the naked eye the Large Magellanic Cloud (LMC) appears like a detached cloud of the Milky Way. The LMC is a barred irregular galaxy of type Ir III-IV, located at a distance of 50 kpc, that has $M_V$ = -18.5. [The absence of a nucleus rules out the classification Sm III-IV (van den Bergh 1998e, p. 27)]. A critical review of various recent distance determinations to the LMC is given by Walker (1999). Because of its proximity even small telescopes can study individual stars that are quite faint. An excellent recent review on the Clouds of Magellan is by Westerlund (1997). For spectacular narrow-band photographs of the LMC in the light of Hα the reader is referred to Davies, Elliot & Meaburn (1976).

### 13.2    Star clusters

Gascoigne & Kron (1952) unexpectedly discovered that the luminous clusters in the Magellanic Clouds fell into two classes. They correctly assumed that red clusters like NGC 1846 were globular clusters. A listing of the 13 genuine globular clusters in the LMC is given by Suntzeff (1992). Surprisingly, Gascoigne & Kron found a second class of clusters in the Large Magellanic Cloud, of which NGC 1866 is the prototype, which are quite luminous and that have blue integrated colors. Such luminous open star clusters are common to the Magellanic Clouds, but appear to be quite rare in the Milky Way system. Van den Bergh & Lafontaine (1984) suggested that this difference was due to the fact that the mass spectrum of open cluster in the Magellanic Clouds was more heavily weighted towards large masses, than it is for open clusters in the Galaxy. An even more extreme overweighting towards massive clusters appears to occur (Fritze-von Alvensleben 1999) in the colliding galaxies NGC 4038/39 (= the Antennae). If this preference for massive cluster formation is due to strong shocks, then an even more extreme overweighting of massive clusters might have occurred during the era when globular clusters formed. Vesperini (1998) shows that such an initial overweighting of massive clusters will persist to the present day (as it does among Galactic globulars), despite destruction of ~ 50% of the original cluster population by relaxation and disk shocking.

### 13.3    History of star and cluster formation

Da Costa (1991) has shown that the rate of cluster formation in the LMC showed an initial peak > 13 Gyr ago, which was followed by "dark ages" that lasted for ~ 9 Gyr, during which only a single cluster was formed. Then, ~ 4 Gyr ago, a violent burst of cluster formation began, which has lasted to the present day. A burst of star formation also started in the Large Cloud 3-4 Gyr ago (Butcher 1977). Perhaps the LMC is the nearest example of the type of object that Babul & Ferguson (1996) have dubbed "boojums", i.e. **b**lue **o**bjects **o**bserved **j**ust **u**ndergoing **m**oderate **s**tarbursts.

Van den Bergh (1998d) has suggested that the increase in the rate of cluster formation ~ 4 Gyr ago was much greater than the corresponding increase in the rate of star formation. The magnitude of this difference is still poorly determined because the history of star formation in the Large Cloud is not yet well constrained by the observations. One might speculate that an increase in strong shocks initiated the recent bursts of cluster formation in the LMC. It should, however, be noted that tidal interactions between the LMC and SMC cannot be the cause of the burst of cluster formation in the LMC. This is so because no corresponding increase in the rate of cluster formation is seen in the SMC. [This difference is significant because one would expect tidal effects of the LMC on the SMC to be stronger than those of the SMC on the LMC.]

Color-magnitude diagrams of globular clusters in the LMC, derived from observations with the Hubble Space Telescope, have been published by Olsen et al. (1998) and by Johnson et al. (1998). These observations show that the LMC clusters are very old, and have ages comparable to those of the oldest Galactic globulars. The radial velocities of LMC globulars appear to show that they belong to a disk system (Schommer et al. 1992). As has already been noted in §12, it is puzzling that the disk of LMC appears to be older than the halo of M 33, even though these two galaxies have comparable luminosities. Bica et al. (1996) have shown that the distribution on the sky of oldest clusters in the LMC has a major axis that is ~ 3 times larger than that for the youngest clusters. Furthermore, Kinman et al. (1991) have drawn attention to the fact that the distribution of the RR Lyrae stars in the LMC may be represented by an exponential with a scale-length of 2.6 kpc, whereas the photometric scale-length for all stars in the Large Magellanic Cloud is only 1.5 kpc (Bothun & Thompson 1988). A similar effect is seen in most Local Group galaxies for which such information is available. In other words galaxies appear to shrink with time. The details of the recent evolution of the Large Cloud are most easily studied using the distribution of Cepheids of differing periods. Such observations show that star formation was more strongly concentrated in the LMC Bar ~ 50 Myr ago than it is at the present time. The absence of really convincing evidence for star and cluster formation associated with a bow-shock suggests that the LMC may presently moving through a region of the outer Galactic halo which is essentially free of interstellar material that is kinematically tied to the Galaxy.

13.4    Interactions between the Magellanic Clouds

The LMC and SMC are separated on the sky by only 20°.7 (~ 19 kpc). The enormous "Magellanic Stream" (Mathewson, Cleary & Murray 1974) represents material that was drawn from these galaxies during an encounter that took place

~ 1.5 Gyr ago.  A second, smaller, and more recent, tidal feature is the "Bridge"

between the LMC and the SMC, that appears to represent material drawn from the

SMC during an interaction that took place only  ~ 0.2 Gyr ago (Gardiner &

Noguchi 1996).  The "Wing" of the SMC is a region of active star formation that

links the Small Cloud to the Bridge.  The tidal interaction between the Clouds

also produced a small leading arm which was first seen by Mathewson & Ford

(1984), and subsequently discussed in detail by Putman et al. (1998).

Observations of individual Cepheids in the SMC (Mathewson, Ford &

Visvanathan 1986, 1988) show that the main body of the SMC has suffered extensive tidal damage, and is now greatly elongated along the line-of-sight.

## 14.     The Small Magellanic Cloud

The SMC is an irregular dwarf galaxy of type Ir IV-V, that has a low

metallicity, and a high mass fraction remaining in gaseous form.  These

characteristics suggest that the Small Cloud is, from an evolutionary point of

view, a more primitive and less evolved galaxy than the Large Cloud.  The fact

that the SMC contains only a single globular cluster (NGC 121) also suggests that

star formation in this object may have ramped up rather slowly.  Reviews on the

Small Cloud have been published by Hodge & Wright (1977) and by Westerlund

(1997).  Various lines of evidence suggest that the distance to the SMC is  ~ 59

kpc.  Some uncertainty is, however, introduced by the fact that the RR Lyrae stars yield a distance modulus to the SMC that is ~ 0.2 mag smaller than that derived from Cepheids. Part of this apparent difference may be due to the fact that the mean Cepheid distance includes Cepheids that are situated in the recently-formed tidal arm behind the SMC. The Wing of the SMC has a distance modulus (Caldwell & Coulson 1986) that appears to be ~ 0.3 mag smaller than that of the main body of the Small Cloud. This observation supports the view that the Wing is part of the Bridge structure linking the SMC (D = 59 kpc) to the LMC (D = 50 kpc). Van den Bergh (1981) has found that the color-magnitude diagram for the (integrated) colors of SMC star clusters exhibits a less-pronounced color gap than does a similar diagram for LMC clusters. The most straightforward explanation for this observation is that the Small Cloud did not experience a major hiatus in its rate of cluster formation, i.e. the SMC never had an extended period ("dark ages") with little or no cluster formation. Some insight into the evolution of the SMC between 1 Gyr and 10 Gyr ago is provided by the distribution of carbon stars (see Figure 18), which has been studied by Morgan & Hatzidimitriou (1995). Hardy,

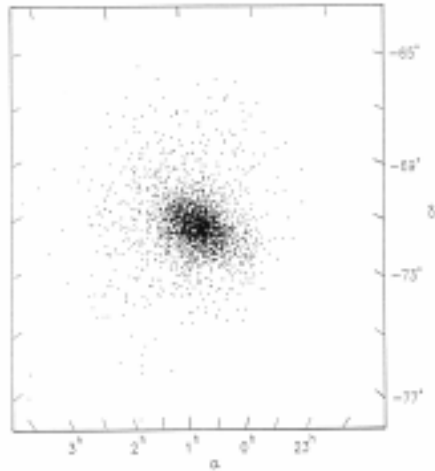

Figure 18    Distribution of carbon stars in the SMC.  The isopleths of the core of the Small Cloud are seen to resemble an E4 galaxy, whereas the outer carbon stars appear to have a more nearly spherical distribution.

Suntzeff & Azzopardi (1989) have used the velocity dispersion of these carbon stars to derive a mass of ~ 5 x $10^8$ M$_\odot$ for the SMC.  Gardiner & Hatzidimitriou (1992) estimate that the oldest SMC population component (which includes RR Lyrae stars), accounts for ~ 7% of the total stellar population in the outer regions of the Small Cloud.

## 15.    The dwarf elliptical M 32 (= NGC 221)

M 32 is the only true elliptical in the Local Group.  The luminosity profile of this object can be represented by an $R^{\frac{1}{4}}$ law, whereas the radial profiles of most spheroidal/dwarf spheroidal galaxies are best represented by an exponential disk. Other important differences between dwarf spheroidals and dwarf ellipticals are discussed by Wirth & Gallagher (1984), and by Kormendy (1985).  M 32 is a close companion of the Andromeda galaxy. The projected separation between the centers of these two objects is only 24′ (5.3 kpc). It was first suggested by Schwarzschild (1954) that the tides produced by M 32 were responsible for the distortion of the spiral structure in, and the warping of, the disk of M 31. Cepa & Beckmann (1988) find that the orbital period of M 32 around M 31 is ~ $8 \times 10^8$ yr. The fact that the orbit of M 32 is retrograde strongly suggests that this object was once captured by the massive (proto) Andromeda galaxy.

Faber (1973) noted that the CN and Mg absorption in M 32 is stronger than it is for most galaxies of similar luminosity. This suggests that M 32 was once a more luminous object that was subsequently tidally stripped of its outer envelope by M 31. The fact that M 32 has no globular clusters, whereas ~ 20 or so are expected, might be due to tidal stripping of the outermost clusters. Such stripped clusters (which would, on average, be more metal-poor than those associated with M 31), are probably still orbiting in the M 31 potential well. The innermost globulars of M 32 might have been dragged into the (very luminous) nucleus of this galaxy by tidal friction.

The semi-stellar nucleus of M 32 (= BD+40° 147) has a maximum rotational V(max) = 55 km s$^{-1}$, and a velocity dispersion σ = 92 km s$^{-1}$. By combining spectroscopic and photometric observations, van der Marel et al.

(1998) find that the nucleus contains a black hole with a mass $M_\bullet = (3.4 \pm 0.7)$ x $10^6 \, M_\odot$. This value represents ~ 0.3% of the total mass of M 32. This is close to the 0.6% value that Merritt (1998) finds for the mean $M_\bullet / M$(spheroid) ratio in galaxies of types E, S0 and Sa.

Magorrian & Tremaine (1999) have pointed out that the tidal disruption of individual stars on radial orbits will produce fuel that can be fed into this central black hole inducing a luminous "flare" that could last for between a few months and a year. They calculate a luminosity of $-17.4 < M_V \text{(max)} < -13.4$ for such a flare. Magorrian & Tremaine estimate a flare frequency of ~ $10^{-4} \, \text{yr}^{-1}$ for the nucleus of M 32. The stellar population of M 32 is dominated by moderately metal-poor red giants. Recent authors have expressed a wide range of opinions on the evolutionary history of M 32. In particular there is still no agreement on whether, or not, this galaxy contains an intermediate-age population component. If we do not even understand the evolutionary history of the nearest elliptical, then some surprises may still be in store for us regarding the history of more distant elliptical galaxies.

## 16.     Local Group Irregular galaxies

### 16.1     The barred irregular NGC 6822

NGC 6822 is a barred irregular with a luminosity $M_V = -16.0$ and of type Ir IV-V.  The distance to NGC 6822 was first determined by Hubble (1925).  He wrote that it was "[A] curiously faithful copy of the [C]louds, but removed to a vastly greater distance".  Reading his papers one has the impression that he was awed by the discovery that the stars in such a distant system seemed to obey the same relationships as nearby ones.  NGC 6822 is far removed from any other members of the LG.  This may account for both the fact that it is (1) an irregular galaxy, and (2) embedded in an enormous flattened hydrogen envelope (Roberts 1972).  This envelope has dimensions of 40′ x 80′ (6 x 12 kpc), compared to optical Holmberg dimensions of 20′ x 20′.  The 21-cm rotation center of this galaxy lies only 61″ ± 20″ south of its optical centroid; i.e. the optical galaxy probably formed very close to the bottom of the NGC 6822 potential well.

### 16.2     The starburst galaxy IC 10

The true nature of IC 10 was first recognized Mayall (1935).  Because of the rather chaotic morphology of this Ir IV:  galaxy Hubble (1936, p. 147) called it "one of the most curious objects in the sky".  The distance to this system is not very well determined due to high (and variable) absorption over its surface. Sakai, Madore & Freedman (1999) obtain distances in the range 500 - 660 kpc from Cepheids, and from the tip of the red giant branch of old evolved stars.

Taken at face value this result suggests that IC 10 is located in front of the M 31

subgroup (see Figure 19), which is centered at a distance of ~ 760 kpc.  If this

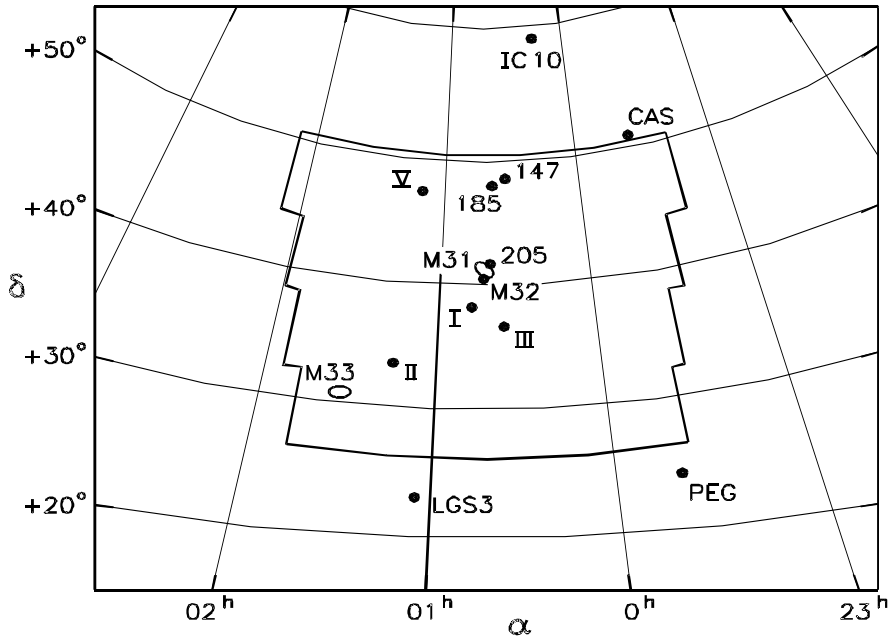

**Figure 19:**    The distribution of galaxies in the Andromeda subgroup of the Local Group.  Note that the dwarf galaxies near the center of this subgroup are all of early type.  The Ir dwarf IC 10 may be located in front of the Andromeda subgroup.  The Pisces system (= LGS 3), which is of type dSph/Ir, is seen

to be located near the outer edge of this subgroup.

conclusion is correct then IC 10 is a free-floating members of the Local Group.  In

that case the LMC + SMC are the only remaining exception to the rule that

irregular galaxies do not occur in LG subclusters.  In this connection it is noted

that (uncertain) orbit calculations by Byrd et al. (1994) suggest that the LMC +

SMC may have been captured into the Galactic subgroup ~ 6 Gyr ago.

Massey & Armandroff (1995) find that IC 10 contains ~ 5 WR stars kpc$^{-2}$, compared to 2 WR stars kpc$^{-2}$ in the very actively star forming LMC. This suggests that IC 10 is a mild example of a starburst galaxy. The nearest active starburst galaxy is IC 1569, which is located at a distance of 2.2 Mpc. IC 10 resembles IC 1569, in that it is also embedded in an amazingly complex filamentary structure that extends to at least 1 kpc from the galaxy's center. Huchtmeier (1979) finds that IC 10 is located within a centrally concentrated H I envelope with dimensions of 62$'$ x 80$'$, which is an order of magnitude greater than the 5.5$'$ x 7.0$'$ dimensions of the optical core of this object. IC 10 therefore belongs to the same class of object as NGC 6822 and IC 1613, which are also embedded in huge hydrogen envelopes. The core of this gas mass (which coincides with the small optical galaxy) rotates in the opposite sense from the outer envelope.

### 16.3    The dwarf irregular IC 1613

The fact that the Ir V dwarf IC 1613, which has $M_V$ = -15.3, was already known in 1906 demonstrates that our sample of Local Group members (except near the plane of the Milky Way) must be complete down to $M_V$ ~ -15. Color-magnitude diagrams show that this object is presently forming stars, and contains both blue and red supergiants that range up to $M_V$ ~ -7.5 in luminosity. Baade (1963, p. 231) has drawn attention to the remarkable fact that IC 1613 appears to contain no star cluster, even though it as actively forming stars. From observations with the <u>Hubble</u> <u>Space</u> <u>Telescope</u>, Hodge & Magnier (Hodge 1998) find that the rate of <u>cluster</u> formation, per unit of star formation, must be at least 600 times lower in IC 1613, than in is the LMC. The reason for this difference is not known. Perhaps, strong shocks (which might be rare in this small quiescent galaxy) play a role in cluster formation. 21-cm observations (Roberts 1972) show that IC 1613 is also embedded in a huge gas envelope. Perhaps such giant gas envelopes are a characteristic of Ir galaxies that are not bound in subclusters, where encounters with other galaxies are frequent.

### 16.4    Other dwarf irregulars

The following faint irregular galaxies are members of the Local Group: Pegasus (= DDO 216), Pisces (= LGS 3), Phoenix, and Leo A (= DDO 69). Possible LG members, that are situated near the outer edge of the Local Group are Aquarius (= DDO 210 = AqrDIG), and SagDIG. These galaxies exhibit a wide range in star forming histories.

Pegasus mainly formed stars 2-4 Gyr ago, although an older population with an age ~ 8 Gyr may also be present. Even at the peak of its star forming activity Pegasus only reached $M_V$ ~ -14 (Tolstoy 1999), so that it never was a "boojum". The older population in Pegasus is more widely dispersed over the surface of this galaxy, than are stars that formed more recently.

The distance to Aquarius is not well established, and no variables have yet been identified in it.  The lack of H II regions demonstrates that the present rate of star formation in this object is low.  It would be interesting to use the Hubble Space Telescope to search for RR Lyrae stars to see if Aqr contains a population component older than ~ 10 Gyr.

Little is known about the star forming history of the Sagittarius dwarf irregular galaxy (SagDIG).  The carbon stars in this object (Cook 1987) suggest that there may have been bursts of star formation between ~ 1 Gyr and ~ 10 Gyr ago.  The fact that the young blue stars in SagDIG are more centrally concentrated than the old carbon stars indicates that this object may have contracted over time. The motions of the gas in SagDIG appear chaotic (Lo, Sargent & Young 1993).

The Pisces dwarf (LGS 3) has a morphology that is intermediate between those of dSph and Ir galaxies.  This suggests that it may be regarded as a dwarf spheroidal that has not yet exhausted its gas supply.  Inspection of Figure 19 shows that LGS 3 is probably an outlying member of the M 31 subgroup of the Local Group.

The Phoenix dwarf is, like LGS 3, an object of type dIr/dSph.  Martínez-Delgado, Gallart & Aparicio (1998) find that the color-magnitude diagram of this galaxy exhibits an extended blue horizontal branch.  This result suggests that Phoenix contains an old stellar population, that might include RR Lyrae variables. It would be important (but difficult) to measure the radial velocity of the brightest stars in this galaxy. This would allow one to tell if the H I clouds seen in this direction are physically associated with Phoenix.

Leo A, which has $M_V = -11.5$, and is of type Ir V, was discovered by Zwicky (1942). It is the most beautiful of the faint dwarf irregulars in the LG. From its color-magnitude diagram Tolstoy (1999) has concluded that the bulk of the stellar population in Leo A is younger than 2 Gyr, and that a major episode of star formation took place 0.9-1.5 Gyr ago. It is not yet clear if Leo A contains any old stars. In this connection a search for RR Lyrae stars would clearly be of great importance. Radio observations by Young & Lo (1996) show that the 8′ x 14′ hydrogen envelope of Leo A is significantly larger than its 4′ x 7′ optical core.

## 17.    Spheroidal galaxies

Wirth & Gallagher (1984), and Kormendy (1985), have shown that early-type galaxies belong to two quite distinct form families:  ellipticals and spheroidals. For the sake of convenience we shall distinguish between relatively bright spheroidals (NGC 147, NGC 185 and NGC 205) and the fainter dwarf spheroidals, of which the Sculptor and Fornax systems are the prototypes. It should, however, be emphasized that the spheroidals and dwarf spheroidals belong to the same morphological class. All three LG spheroidals are members of the Andromeda subgroup of the Local Group (see Figure 19). In all of these spheroidals rotation is negligibly small, which shows that their observed flattenings are mainly due to velocity anisotropies.

### 17.1    NGC 205

This spheroidal galaxy has $M_V = -16.4$, which is similar to that of the dwarf elliptical M 32, for which $M_V = -16.5$. However, NGC 205 has a much more extended structure than does M 32. Furthermore M 32 rotates, whereas NGC 205 does not (Bender, Paquet & Nieto 1991). Carter & Sadler (1990) find that $V(max)/\sigma = 0.2$, in which $\sigma$ is the velocity dispersion. This demonstrates that NGC 205 is mainly flattened by its velocity anisotropy. Deep exposures show that the outer isophotes of NGC 205, which is located at only 37′ (8 kpc) from M 31, are distorted by the tides induced by the Andromeda galaxy. Dust patches, and some regions of low-level star formation, are embedded within the main body of NGC 205. The light of the nucleus of NGC 205, which has $L_V \sim 4 \times 10^4 \, L_\odot$, appears to be dominated by stars with ages of $1\text{-}5 \times 10^8$ yr. If a black hole is present in this nucleus, its mass must be $< 1 \times 10^5 \, M_\odot$ (Jones et al. 1996). Welch, Sage & Mitchell (1998) show that the amount of gas that is present in NGC 205 is an order of magnitude lower than the gas mass that would have been ejected by evolving stars during their lifetimes. This suggests that a great deal of gas must have been ejected by supernovae and stellar winds, or swept out during a passage through the fundamental plane of M 31.  The fact that the gaseous disk of this galaxy is rotating, whereas the parent galaxy is not, shows that some high angular momentum gas must have been captured by NGC 205 during the course of its evolution.

### 17.2    NGC 185

NGC 185, and its companion NGC 147, appear to be slightly less distant than M 31.  Da Costa & Mould (1988) find that the globular clusters associated with NGC 185 have  <[Fe/H]> = -1.65, which is significantly lower than the value <[Fe/H]> = -1.23 that Lee et al. (1993c) find for the red giants in this galaxy. Similarly the globular clusters in NGC 205 are metal-poor compared to the stars in the main body of this galaxy.  This indicates that the globulars in dwarf spheroidals were formed early in the history of these spheroidal galaxies. Kinematical studies by Bender, Paquet & Nieto (1991) show that NGC 185 has a rotational velocity of $1.2 \pm 1.1$ km s$^{-1}$.  This shows that rotation does not provide a major contribution to the observed flattening of NGC 185.  Some star formation is still taking place near the center of this galaxy.  The fact that the gas has similar kinematics to that of the stars (Young & Lo 1997) suggests that it may have originated in stellar winds.

17.3    NGC 147

The spheroidal NGC 147 was first resolved by Baade (1944b) in his pioneering study of stellar populations.  It differs from its more luminous companion NGC 185 in that it has a lower surface brightness, and because it does not exhibit any dust absorption.  Bender, Paquet & Nieto (1991) find that V(rot) = 6.6 km s$^{-1}$, which is much smaller than its velocity dispersion of $\sigma$ = 23 km s$^{-1}$. This shows that the flattening of NGC 147 must be mainly due to its anisotropic velocity dispersion.  It is not clear (Hodge 1976) whether the object near the center of NGC 147 is its nucleus, or if it is a globular cluster.  The discovery of RR Lyrae stars in NGC 147 (Saha, Hoessel & Mossman 1990) provides prima facie evidence for the existence of a very old population with age > 10 Gyr in this galaxy.  From the internal velocity dispersions of, and velocity difference between NGC 147 and NGC 185, van den Bergh (1998c) has shown that these two galaxies (which have a projected separation of only 11 kpc) form a physical binary system.

## 18.    The brightest dwarf spheroidals

18.1    The Sagittarius dwarf spheroidal

The Sagittarius dwarf spheroidal galaxy, which appears to have been partly disrupted by Galactic tides, was discovered serendipitously by Ibata, Gilmore & Irwin (1994).  It had previously eluded us because it is projected on the rich star fields associated with the Galactic bulge. Mateo, Olszewski & Morrison (1999) appear to have detected stars associated with this object at a distance of $34°$ from its center. This suggests that the outer parts of this galaxy might be thought of as a star-stream, rather than as an extension of its inner isophotes. At even larger angular distances the globular clusters Palomar 2 and Palomar 12 may represent objects that were tidally stripped from Sagittarius (Irwin 1999). Ibata et al. noted that the very luminous globular cluster NGC 6715 (= M 54) is located close to the center of Sagittarius, and might therefore be its nucleus. A possible argument against this hypothesis is, however, that the fraction of spheroidal/dwarf spheroidal galaxies that has nuclei drops precipitously with decreasing parent galaxy luminosity, from ~ 100% at $M_V$ = -17, to ~ 10% at $M_V$ = -12 (van den Bergh 1986). However, this argument loses some of its force if Sagittarius, which now has a very uncertain luminosity of $M_V$ = -13.8, was once much more luminous than it is now. Ibata et al. also found that the globular clusters Arp 2, Terzan 7 and Terzan 8 are grouped around, and probably associated with, the Sagittarius system. Van den Bergh (1998d) has noted that the luminosity distribution of the globulars associated with Sagittarius (one very luminous, and four faint) appears to differ significantly from the Gaussian luminosity distribution of all Galactic globulars, but resembles that in the outermost Galactic halo at $R_{GC}$ > 80 kpc. This suggests that Sagittarius might possibly be a Searle-Zinn (1978) fragment formed in the outer halo that was later captured by the inner Galaxy, after an encounter with the Magellanic Clouds.

This argument is supported by the observation that both the outer halo and Sagittarius contain "young" globulars.  Such objects are found to be absent from the region of the Galactic halo with $R_{GC} < 15$ kpc.

According to Bellazini, Ferraro & Buonanno (1999) star formation in the Sagittarius galaxy started early and ended abruptly  ~ 8 Gyr ago; probably at the time that the gas in this object was suddenly depleted.  Montegriffo et al. (1998) find that Arp 2 is  ~ 4 Gyr younger than the globular cluster M 54, while Terzan 7 is 6-7 Gyr younger. The metallicity of Terzan 7 is an order of magnitude higher than it is in the other globulars associated with this galaxy.  The dominant stellar population in the Sagittarius dwarf appears to have an age and metallicity very similar to that of Terzan 7.  It is not yet clear why dwarf spheroidal galaxies were able to form globular clusters much later than the Milky Way system itself.

### 18.2     The Fornax dwarf spheroidal

The Fornax system was discovered by Shapley (1939), who later noted (Shapley 1943, pp. 142-142) that such objects fell outside the Hubble (1936) classification system, and that there seemed to be no need for their existence! [We are presently faced with a similar conundrum because there appears to be no logical need for the existence of  "oversized dSph" galaxies, such as F8D1 in the M 81 group (Caldwell et al. 1998).]

Because of its large size, and low surface brightness, the integrated magnitude of the Fornax system is difficult to determine. Absolute magnitudes in the recent literature range from $M_V$ = -13.9 to $M_V$ = -13.2. As a result both the mass-to-light ratio of Fornax, and its specific cluster frequency S (Harris & van den Bergh 1981), which is the number of globular clusters per $M_V$ = -15 of parent galaxy luminosity, remain uncertain by a factor of ~ 2. Five globular clusters are associated with the Fornax galaxy, of which NGC 1049 is the most luminous. The specific globular cluster frequency of Fornax is therefore $13 \le S \le 26$. This frequency, which is higher than that in most early-type galaxies (van den Bergh 1998f), places significant constraints on theories of globular cluster formation. The Fornax globulars are unusual because their horizontal branches are, at a given metallicity, redder than those of the majority of their Galactic counterparts. They might therefore be regarded as extreme examples of the NGC 7006 "second parameter" effect. Fornax clusters Nos. 1, 2, 3 and 5 are found to have ages that are identical to within ~ 1 Gyr. Furthermore this age appears to be indistinguishable from that of the old, metal-poor Galactic globular M 92. However, Marconi et al. (1999) find that Fornax No. 4 is significantly younger than the other globulars associated with the Fornax dwarf. The mean luminosity of the Fornax globulars is $<M_V>$ = -7.1 ± 0.5. This value is, within its error bars, indistinguishable from $<M_V>$ = -7.4 in giant galaxies (Harris 1991). This shows that the luminosity function of globular clusters is insensitive to the luminosities of their parent galaxies.

The color-magnitude diagram for the main body of Fornax has been discussed by Stetson, Hesser & Smecker-Hane (1998), who find stars with ages between 1 Gyr and 10 Gyr.  The stellar surface brightness distribution in Fornax is asymmetrical due to star formation during the most recent phase of its evolution.  Stetson et al. conclude that the oldest population, which includes RR Lyrae stars, is the most widely dispersed.  Intermediate-age stars, which include red clump giants, are more centrally concentrated.  Finally the youngest stellar population appears to have a more flattened distribution.  Mateo et al. (1991) find that the central density in Fornax is $0.07 \pm 0.03$ M$_\odot$ pc$^{-3}$.  This value is comparable to that found in gas-rich dwarf irregulars, but smaller than that in most other dwarf spheroidals.  Mateo et al. find no evidence for a significant rotation of the Fornax dwarf around its minor axis.  In this respect Fornax resembles the more luminous spheroidal companions of M 31.

## 19.    Faint dwarf spheroidal galaxies

At the present time (see Table 1) 18 LG members fainter than $M_V = -14.0$ are known.  Of these objects 13 (72%) are dSph, two (11%) are dIr, two (11%) are of the transitional type dSph/dIr, and one (6%) is of unknown morphological type. These data show that dwarf spheroidal galaxies are the most frequent type of galactic system in the LG, and hence (presumably) in the entire Universe.

### 19.1 The Andromeda subgroup

Six dwarf spheroidal galaxies fainter than $M_V = -13.0$ are presently known to be located in the Andromeda subgroup of the Local Group (See Figure 19). The fact that only one of these objects is fainter than $M_V = -10.0$ suggests that the sample of dSph galaxies in this subgroup may still be quite incomplete. The first three of these dwarfs, And I, And II and And III, were detected as faint smudges on IIIaJ plates obtained with the Palomar 1.2-m Schmidt telescope (van den Bergh 1972a). Subsequent observations with the Hale 5-m. reflector (van den Bergh 1972b) showed that And III resolved into stars at about the same brightness level as the giants of Population II in M 31 itself. Subsequently (van den Bergh 1974) And I and And II were also shown to resolve at the same magnitude. A color-magnitude diagram obtained by Da Costa et al. (1996) showed that the bulk of the stellar population in And I has an age of $\sim 10$ Gyr. The fact that And II contains one or two C stars (Aaronson et al. 1985) shows that this object contains some intermediate-age objects. Furthermore Armandroff et al. (1993) found that $10 \pm 10\%$ of the luminosity of And III is contributed by stars with ages of 3-10 Gyr. A fourth dwarf spheroidal companion to M 31, which they dubbed And V, was found by Armandroff, Davies & Jacoby (1998). The additional dSph galaxies And VI (= Pegasus II) and And VII (= Cassiopeia) have been observed by Grebel & Guhathakurta (1999). Since the digitization of the second <u>Palomar Sky Survey</u> is not yet complete, additional faint members of the Andromeda subgroup of the LG may remain to be discovered. No systematic differences (van den Bergh 2000) have been found between the dSph galaxies in the Andromeda and Galactic subgroups of the Local Group. In particular many of the dwarf spheroidals in these two groups seem to have continued to form stars for a significant fraction of a Hubble time.

### 19.2    Faint companions to the Milky Way

### 9.2.1    Leo I (= Regulus)

The fact that this dSph galaxy is located at a distance of $250 \pm 15$ kpc makes it uncertain whether it is, in fact, a member of the Galactic subgroup of the LG. This question is of critical importance because Leo I has a large velocity deviation from the mean $V_r$ versus $\cos \theta$ relation for LG members (see Figure 2). If it is a member, then the mass of the Galactic subgroup is $12 \times 10^{11}$ $M_\odot$. However, if it is not then the mass of the Galactic subgroup is only $4 \times 10^{11}$ $M_\odot$. Aaronson, Olszewski & Hodge (1983) have shown that Leo I contains both carbon stars and ~ 50 asymptotic branch stars above the tip of the red giant branch. This shows that Leo I has a significant intermediate-age population. More recently Gallart et al. (1998, 1999) have shown that the bulk of the stellar population in Leo I has an age of only 2-6 Gyr. Caputo et al. (1999) find no evidence for the presence of RR Lyrae stars in this object. Leo I and Leo A might turn out to be the only LG members in which there was no burst of star formation ~ 15 Gyr ago.

### 19.2.2  Carina

Mateo et al. (1993) have obtained $M/L_V \approx 23$ (in solar units), which suggests that most of the mass of the Carina system is probably in the form of dark matter. Mould & Aaronson (1983) found a mean age of $7.5 \pm 1.5$ Gyr for the stars in Carina. The fact that RR Lyrae variables occur in this object shows that at least a few percent of the stellar population in this galaxy must be old. More recently Hurley-Keller, Mateo & Nemec (1998) have derived a rather complex evolutionary history for this galaxy. They find that it underwent significant pulses of star formation ~ 15 Gyr, ~ 7 Gyr and ~ 3 Gyr ago. Only 10-15% of this star formation took place during the initial burst. During the main burst, that took place ~ 7 Gyr ago, Carina would have had $M_V$ ~ -16. Furthermore these authors conclude that ~ 30% of the stars in Carina formed during the most recent burst that occurred ~ 3 Gyr ago. The fact that the mean period of the RR Lyrae stars $<P_{ab}> = 0.62$, which is intermediate between that of Oosterhoff's types I and II, shows that the early evolutionary history of Carina must have differed from that of Galactic globular clusters. The same conclusion is found to hold for the RR Lyrae stars in other dwarf spheroidals, and for those in the Clouds of Magellan.

### 19.2.3 Sculptor

The Sculptor system was discovered by Shapley (1938). Hurley-Keller, Mateo & Grebel (1999) have mapped most of this object in B and R. They find that the (presumably very old) blue horizontal branch stars occur at all radii, whereas the younger red horizontal branch stars are mainly located near the center of this galaxy. This shows that Sculptor is another example of a dwarf galaxy that shrank as it aged. Demers & Battinelli (1998) have found nearly 200 blue stars brighter than the main sequence turn-off. These young stars may be associated with the $\geq 3 \times 10^4$ $M_\odot$ of neutral hydrogen gas that Carignan et al. (1998) have found to be associated with this galaxy.

### 19.2.4 Draco

The Draco system was discovered by Wilson (1955). Baade & Swope (1961) found that this object has a globular cluster-like color-magnitude diagram. Hubble Space Telescope observations by Grillmair et al. (1998) show no evidence for multiple main sequence turnoffs. This suggests that most of the star formation in Draco took place during a single initial burst. However, the discovery of three C stars by Aaronson (1983) shows that the Draco system does contain a few intermediate-age stars. A comparison between the color-magnitude diagram of Draco, and that of the old metal-poor Galactic globular clusters, shows that the Draco dSph galaxy is 1.6 ± 2.5 Gyr older than M 92. The fact that the Draco system is both very metal-poor, and has a red horizontal branch, demonstrates that the second parameter effect cannot (entirely) be an age effect. From high dispersion spectra of four stars Shetrone, Bolte & Stetson (1998) have found that the slope of the [Ba/Fe] versus Fe relation in Draco is steeper than that found in Galactic globular clusters and in the Galactic halo. If confirmed, this result would militate against the hypothesis that most of the Galactic halo was built up from disintegrated dwarf spheroidal-like objects. From its structural parameters and radial velocity dispersion the mass-to-light ratio in Draco is found to be ~ 90 in solar units. This shows that most of the mass of this object is in the form of dark matter.

### 19.2.5  Tucanae

This dwarf galaxy, which is located at a distance of 870 ± 60 kpc, is the only known dwarf spheroidal in the Local Group that is <u>not</u> associated with either the Galactic, or Andromeda, subgroups. This shows that such objects can form outside the deep potential wells surrounding major galaxies. According to Da Costa (1998), Tucanae contains RR Lyrae variables and bright red and blue horizontal branch stars.  These red horizontal branch stars might belong to a stellar population with an age of 3-4 Gyr.

### 19.2.6  Leo II

The dwarf spheroidal Leo II was discovered by Harrington & Wilson (1950) on the first plates obtained with the Palomar 1.2-m Schmidt telescope. Demers & Irwin (1993) found that the horizontal branch of this object is heavily populated to the red of the RR Lyrae gap, but also contains a few (possibly old) blue horizontal branch stars.  Mighel & Rich (1996 a,b) find that star formation in Leo II started $14 \pm 1$ Gyr ago, and reached its peak $9 \pm 1$ Gyr ago.  The rate of star formation in Leo II appears to have been negligibly small during the last  $\sim 7$ Gyr.  From the velocity dispersion of 31 giants Vogt et al. (1995) found this galaxy to have $M/L_V \sim 12$ in solar units; i.e. most of its mass is in the form of dark matter.

### 19.2.7  Sextans

The Sextans system was discovered by Irwin et al. (1990).  From the velocity dispersion measured by Suntzeff et al. (1993), $M/L_V \sim 50$ in this galaxy. This shows that most of its mass must be in the form of dark matter.  These authors also find a metallicity range of $0.19 \pm 0.02$ dex in Sextans.  This shows that the stars in the Sextans dwarf have a significant range in metallicity.  The rotational velocity of Sextans is found to be < 0.4 km s$^{-1}$.  The presence of significant numbers of RR Lyrae stars shows that star formation in Sextans must have started early.  Mateo, Fischer & Krzeminski (1995) find that ∼ 25% of the population of Sextans has ages of 2-4 Gyr.

### 19.2.8  Ursa Minor

The UMi system was discovered on Palomar Sky Survey plates by Wilson (1955).  From the radial velocity observations of Armandroff, Olszewski & Pryor (1995) the mass-to-light ratio of the Ursa Minor system is found to be $M/L_V = 77 \pm 13$, in solar units.  This shows that the overwhelming majority of the mass in UMi is in invisible form.  The color-magnitude diagram of UMi appears to show no evidence for stars younger than ∼ 15 Gyr.  This suggests that most of the stars in the Ursa Minor system were made during a single burst of star formation.

### 19.3    Common characteristics of dSph galaxies

The dwarf spheroidal galaxies in the Local Group share a number of important characteristics.  With the exception of Dra and UMi, these objects appear to have been able to retain gas for a significant fraction of a Hubble time.  Furthermore most of these galaxies seem to have had a rather complex history of star formation.  With the possible exception of Leo I, star formation in all of these objects seems to have started simultaneously ∼ 15 Gyr ago.  Dwarf spheroidal galaxies resemble the more luminous spheroidals like NGC 147, NGC 185 and NGC 205, in that they exhibit little, if any, evidence for rotation. This shows that dSph galaxies cannot be the descendents of rotating dIr galaxies. A plot of the masses versus the luminosities of dSph galaxies is shown in Figure 20. This

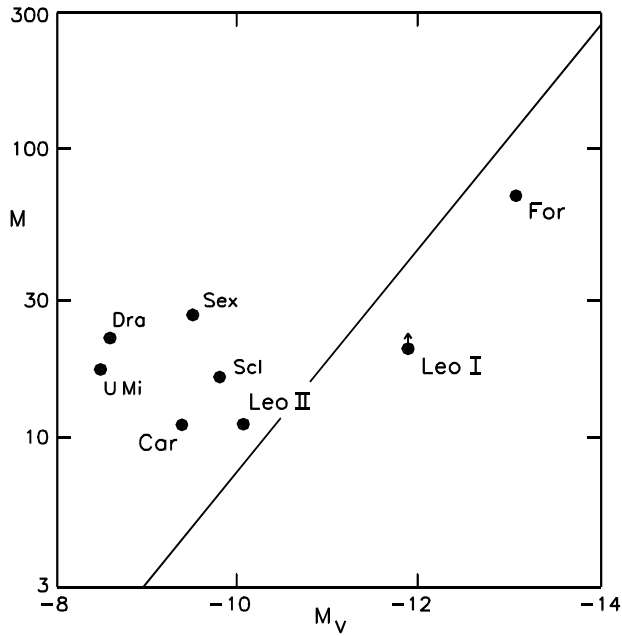

**Figure 20:**    Mass (in units of $10^6$ M$_\odot$) versus M$_V$ for dSph galaxies. The diagonal line is the locus of points with M/L$_V$ = 10. The least luminous galaxies are seen to have the highest mass-to-light ratios. Note absence of objects with M < $10^7$ M$_\odot$.

figure shows that the faintest dwarf spheroidals have the highest mass-to-light ratios. Presumably part of the reason for this is that the least massive dSph galaxies lost a significant fraction of their initial gas mass in supernova driven stellar winds. A remarkable feature of Figure 20 is that no known dSph galaxy has M < 1 x $10^7$ M$_\odot$. It is possible that such objects exist, but have lost all of their baryonic matter, so that they cannot form any stars. It would be interesting

to know how much matter might be hidden in such small dark matter "cannonballs".

The known Local Group galaxies fainter than $M_V = -12.0$ have a combined integrated magnitude $M_V = -13.3$ (van den Bergh 1998b), but contain not a single globular cluster. The total number of clusters expected to be associated with these objects is 0.21S, in which S is the specific globular cluster frequency. From Poisson statistics one finds that there is an a priori probability of finding no globular clusters of 0.81 for S = 1, of 0.35 for S = 5 and of 0.015 for S = 20. It follows that is quite improbable that typical faint dwarf spheroidal galaxies have a high specific frequency, like that observed in the Fornax system.

Dottori, Mirable & Duc (1994) have proposed that some dSph galaxies might originally have formed in tidal tails. Such objects would be expected to have low mass-to-light ratios. This is so because dark matter that is pulled out of giant galaxies is stretched (Kormendy 1998). As a result fragments pulled out of more luminous galaxies will have lower densities than their progenitors. The fact that none of the low-luminosity galaxies in Figure 20 exhibits a low M/L ratio shows that none of these dSph galaxies were formed from tidal tails.

## 20.  Conclusions

Modern research has thoroughly vindicated Hubble's expectation that the Local Group of galaxies could produce new insights into the nature of stellar systems, and that it would also provide stepping stones on the path leading to the more remote reaches of the Universe.  However, even Hubble himself would probably have been surprised by the amount of information that the Local Group has provided on the evolution of galaxies, and on conditions at the time when galaxies first formed.

One of the most interesting results obtained from studies of individual LG members is that most galaxies already started to form stars ~ 15 Gyr ago.  The only possible exceptions to this rule are the (unrelated) galaxies Leo I and Leo A. In other words star formation started with a bang and, at least in early-type galaxies, ended with a whimper.  It is puzzling that even the remote globular cluster NGC 2419, at a distance of ~ 100 kpc from the Galactic center, formed at this early date.  Why did most star creation in the Local Group start simultaneously in different galaxies and in differing environments?

Modern observations confirm Hubble's (1936, p. 125) belief that the Local Group is "a typical, small group of nebulae which is isolated in the general field." In all those LG galaxies for which such information is available, old stars are found to be more widely dispersed than younger stars; i.e. galaxies are observed to shrink with time.  It is not yet clear how this observation is to be reconciled with the observation (Bouwens, Broadhurst & Silk 1998, Pascarelle, Windhorst & Keel 1998) that distant galaxies in the Hubble Deep Field appear to be smaller than the nearby ones that are observed at z ~ 0.  It is also of interest to note that none of the low-luminosity members of the LG appear to have been luminous "boojums" during the last  ~ 10 Gyr.  It would also be of interest to study the possible effects of the "mini-quasars", that probably once existed in the nuclei of M 31, M 32 and the Milky Way, on their environment.

An observation, that is difficult to reconcile with current ideas on galaxy evolution, is that the old LMC globular clusters appear to be located in a disk, while the apparently younger globulars in M 33 seem to belong to a halo population.  Why did the M 33 halo form before the LMC disk, even though the LMC and M 33 have similar masses?  Another important question posed by LG observations is why the LMC remained in an almost dormant stage for  ~ 9  Gyr, before its current burst of star and cluster formation started  ~ 4 Gyr ago.

The fact that the Sagittarius galaxy is presently being disrupted by Galactic tides, and the existence of the Magellanic Stream, and Magellanic Bridge, show that galaxies do not evolve in isolation, and that interactions can play an important role in galaxy evolution.  Other evidence for such interactions is provided by the apparent truncation of M 32, and by the tidal distortions of M 31, the SMC and NGC 205.  Furthermore, the fact that the spheroidal component of the Andromeda galaxy exhibits an $R^{1/4}$ profile hints at merger and violent relaxation early in the history of M 31.  On the other hand the balance of evidence appears to suggest that the bulk of the Galactic halo was not derived from mergers with dwarf spheroidal galaxies, and the disintegration of globular clusters.  The fact that the most luminous LG irregular galaxies are located outside subgroups, and have very large hydrogen envelopes, suggest that such objects preferentially formed in isolation.

In the Realm of the Nebulae (1936, p. 125) Hubble wrote "The nearer nebulae in the general field are far beyond the limits of the [Local] group. Investigation of their stellar contents are so difficult that very little definite information has been gathered…The fact that the [G]alactic system is a member of a group is a very fortunate accident."  In the present review I have tried to show how modern observations of the members of the Local Group have taken advantage of the "very fortunate accident" that the Milky Way system is a member of such a small group of galaxies. This has allowed us to study the stellar content, and evolutionary history, of a representative sample of galaxies in a way that would have been quite impossible if the Galaxy had been an isolated field galaxy.

# REFERENCES

Aaronson, M. 1983, ApJ, 166, L11

Aaronson, M., Gordon, G., Mould, J., Olszewski, E., & Suntzeff, N. 1985, ApJ, 296, L7

Aaronson, M., Olszewski, E.W. & Hodge, P.W. 1983, ApJ, 267, 271

Appleton, P.N., Davies, R.D. & Stephenson, R.J. 1981, MNRAS, 195, 327

Armandroff, T.E. 1998, Colloquium given in Victoria, 1998 May 25

Armandroff, T.E., Da Costa, G.S., Caldwell, N. & Seitzer, P. 1993, AJ, 106, 986

Armandroff, T.E., Davies, J.E. & Jacoby, G.H. 1998, AJ, 116, L287

Armandroff, T.E., Olszewski, E.W. & Pryor, C. 1995, AJ, 110, 2131

Baade, W. 1944a, ApJ, 100, 137

Baade, W. 1944b, ApJ, 100, 147

Baade, W. 1951, Publ. Obs. U. Michigan, 10, 7

Baade, W. 1963, Evolution of Stars and Galaxies, (Cambridge: Harvard University Press)

Baade, W. & Swope, W.W. 1961, AJ, 66, 300

Babcock, H.W. 1939, Lick Obs. Bull. 19, 41, (No. 498)

Babul, A. & Ferguson, H.C. 1996, ApJ, 458, 100

Backer, D.C. & Sramek, R.A. 1982, ApJ, 260, 512

Bahcall, J.N. & Tremaine, S. 1981, ApJ, 244, 805

Barnes, J. 1990, in Dynamics and Interactions of Galaxies, Ed. R. Wielen (Berlin: Springer), p. 186

Bellazzini, M., Ferraro, F.R. & Buonanno R. 1999 MNRAS, in press

Bender, R., Paquet, A. & Nieto, J.-L. 1991, A&A, 246, 349

Berman, B.G. & Suchkov, A.A. 1991, Ap. Space Sci., 184, 169

Bica, E., Claría, J.J., Dottori, H., Santos, J.F.C. & Piatti, A.E. 1996, ApJS, 102, 57

Binney, J. & Tremaine, S. 1987, Galactic Dynamics, (Princeton:  Princeton Univ.
     Press)

Blair, W.P., Kirshner, R.P. & Chevalier, R.A. 1981, ApJ, 247, 879

Bland-Hawthorn, J. & Maloney, P.R. 1998, ApJ, 510, L33

Bland-Hawthorne, J., Veilleux, S., Cecil, G.N., Putman, M.E., Gibson, B.K. &
     Maloney, P.R., 1998, MNRAS, in press

Blitz, L. & Spergel, D.N. 1991, ApJ, 379, 631

Blitz, L., Spergel, D.N., Teuben, P.J., Hartman, D. & Burton, W.B. 1998, astro-
     ph/9803251

Bothun, G.D. & Thompson, I.B. 1988, AJ, 96, 877

Bouwens, R., Broadhurst, T. & Silk, J. 1998, ApJ, 506, 557

Burstein, D., Faber, S.M., Gaskell, C.M. & Krumm, N. 1984, ApJ, 287, 586

Burton, W.B. & Braun, R. 1999, A&A, in press

Butcher, H. 1977, ApJ, 216, 372

Byrd, G., Valtonen, M., McCall, M. & Innanen, K. 1994, AJ, 107, 2055

Caldwell, J.A.R. & Coulson, I.M. 1986, MNRAS, 218, 223

Caldwell, N., Armandroff, T.E., Da Costa, G.S. & Seitzer, P. 1998, AJ, 115, 535

Caldwell, N., Armandroff, T.E., Seitzer, P. & Da Costa, G.S. 1992, AJ, 103, 840

Caldwell, N. & Schommer, R. 1988, in The Extragalactic Distance Scale (= ASP
Conference Series No. 4), Eds. S. van den Bergh and C.J. Pritchet, (San
Francisco:  ASP), p. 77

Caputo, F., Cassisi, S., Castellani, M., Marconi, G. & Santolamazza, P. 1999, AJ,
in press

Carignan, C., Beaulieu, S., Côté, S., Demers, S. & Mateo, M. 1998, AJ, 116, 1690

Carney, B.W., Latham, D.W. & Laird, J.B. 1990, AJ, 99, 572

Carter, D. & Sadler, E.M. 1990, MNRAS, 245, 12p

Cepa, J. & Beckman, J.E. 1988, A&A, 200, 21

Chaboyer, B., Green, E.M. & Liebert, J. 1999, AJ, 117, 1360

Chandar, R. 1998, BAAS, 30, 1336

Chamcham, K. & Hendry, M.A. 1996, MNRAS, 279, 1083

Chappell, D. & Scalo, J. 1997, preprint

Chen, B. 1998, ApJ, 495, L1

Combes, F., Debbasch, F., Friedli, D. & Pfenniger, D. 1990, A&A, 233, 82

Combes, F. & Pfenniger, D. 1997, A&A, 327, 453

Cook, K.H. 1987, Arizona PhD Thesis

Courteau, S. & van den Bergh, S. 1999, AJ, in press

Crane, P.C., Dickel, J.R. & Cowan, J.J. 1992, ApJ, 390, L9

Da Costa, G.S. 1991, in The Magellanic Clouds (= IAU Symposium No. 148),
Eds. R. Haynes and D. Milne, (Dordrecht:  Kluwer), p. 183

Da Costa, G.S. 1998, in Stellar Astrophysics for the Local Group, (Cambridge, Univ. of Cambridge Press), p. 351

Da Costa, G.S., Armandroff, T.E., Caldwell, N. & Seitzer, P. 1996, AJ, 112, 2576

Da Costa, G.S. & Mould, J.R. 1988, ApJ, 334, 159

Dame, T.M., Koper, E., Israel, F.P. & Thaddeus, P. 1993, ApJ, 418, 730

Davies, R.D., Elliot, K.H. & Meaburn, J. 1976, Mem. RAS, 81, 89

Davis, D.S., Bird, C.M., Mushotzky, R.F. & Odewahn, S.C. 1995, ApJ, 440, 48

de Boer, K.S. 1999, in Harmonizing the Cosmic Distance Scale in the Post-Hipparcos Era, Eds. D. Egret and A. Heck, (San Francisco: ASP), in press

de Boer, K.S., Braun, J.-M., Vallenari, A. & Mebold, U. 1998, A&A, 329, L49

Dekel, A. & Silk, J. 1986, ApJ, 303, 39

Demers, S. & Battinelli, P. 1998, AJ, 115, 154

Demers, S. & Irwin, M.J. 1993, MNRAS, 261, 657

Diaconis, P. & Efron, B. 1983, Sci. Am., 248, No. 5, p. 96

Dottori, H., Mirabel, F. & Duc, P.-A. 1994, in Dwarf Galaxies, Eds. G. Meylan and P. Prugniel, (Garching: ESO), p. 393

Dressler, A. & Richstone, D.O. 1988, ApJ, 324, 701

Eckart, A. & Genzel, R. 1997, MNRAS, 284, 576

Edvardsson, B., Andersen, J., Gustafsson, B., Lambert, D.L., Nissen, P.E. & Tomkin, J. 1993, A&A, 275, 1

Eggen, O.J., Lynden-Bell, D. & Sandage, A.R. 1962, ApJ, 136, 748

Einasto, J., Saar, E., Kaasik, A. & Chernin, A.D. 1974, Nature, 252, 111

Faber, S.M. 1973, ApJ, 179, 731

Freeman, K.C. 1999, in Stellar Content of the Local Group, (= IAU Symposium
        No. 192, Eds. P. Whitelock and R.D. Cannon, (San Francisco:  ASP), in
        press

Freedman, W.L. et al. 1994, ApJ, 427, 628

Friel, E.D. 1995, ARAA, 33, 381

Fritze-von Alvensleben, U. 1999, A&A, 342, L25

Fusi Pecci, F., Buonanno, R., Cacciari, C., Corsi, C.E., Djorgovski, S.G.,
        Frederici, L., Ferraro, F.R., Parmeggiano, G. & Rich, R.M. 1996, AJ,
        112, 1461

Fusi Pecci, F., Cacciari, C., Frederici, L. & Pasquali, A. 1993, in The Globular
        Cluster - Galaxy Connection (= ASP Conference Series Vol. 48), Eds.
        G.H. Smith and J.P. Brodie, (San Francisco:  ASP), p. 410

Gallart, C. et al. 1999, ApJ, in press

Gallart, C., Aparicio, A., Freedman, W., Bertelli, G. & Chiosi, C. 1998, in The
        Magellanic Clouds and other Dwarf Galaxies, Eds. T. Richtler and J.M.
        Braun, (Aachen:  Shaker Verlag), p. 147

Gardiner, L.T. & Hatzidimitriou, D. 1992, MNRAS, 257, 195

Gardiner, L.T. & Noguchi, M. 1996, MNRAS, 278, 191

Gascoigne, S.C.B. & Kron, G.E. 1952, PASP, 64, 196

Ghez, A.M., Klein, B.L., Morris, M. & Becklin, E.E. 1998, ApJ, 509, 678

Gordon, K.G. 1969, QJRAS, 10, 293

Gordon, K.D., Hanson, M.M., Clayton, G.C., Rieke, G.H. & Misselt, K.A. 1999, astro-ph/9902043

Grebel, E.K. 1997, Reviews of Modern Astr., 10, 29

Grebel, E.K. & Guhathakurta, P. 1999, in Stellar Content of the Local Group, (= IAU Symposium No. 192), Eds. P. Whitelock and R.D. Cannon, (San Francisco:  ASP), in press

Grillmair, C.J. et al. 1998, AJ, 115, 144

Hardy, E., Suntzeff, N.B. & Azzopardi, M. 1989, ApJ, 344, 210

Harringon, R.G. & Wilson, A.G. 1950, PASP, 62, 118

Harris, W.E. 1991, ARAA, 29, 543

Harris, W.E. et al. 1997, AJ, 114, 1030

Harris, W.E. & van den Bergh, S. 1981, AJ, 86, 1627

Heiles, C. 1987, ApJ, 315, 555

Heisler, J., Tremaine, S. & Bahcall, J.N. 1985, ApJ, 298, 8

Hodge, P.W. 1976, AJ, 81, 25

Hodge, P.W. 1998, Colloquium given in Victoria, 1998 April 21

Hodge, P.W. & Wright, F.W. 1977, The Small Magellanic Cloud, (Seattle:  Univ. of Washington Press)

Hubble, E. 1925, ApJ, 62, 409

Hubble, E. 1936, The Realm of the Nebulae, (New Haven:  Yale University Press)

Huchtmeier, W.K. 1979, A&A, 75, 170

Hurley-Keller, D., Mateo, M. & Grebel, E. 1999, in preparation

Hurley-Keller, D., Mateo, M. & Nemec, J. 1998, astro-ph/9804058

Ibata, R.A, Gilmore, G. & Irwin, M.J. 1994, Nature, 370, 194

Idiart, T.P., de Freitas Pacheco, J.A. & Costa, R.D.D. 1996, AJ , 113, 1169

Innanen, K.A., Kamper, K.W., Papp, K.A. & van den Bergh, S. 1982, ApJ, 254, 515

Irwin, M.J. 1999, in The Stellar Content of the Local Group (= IAU Symposium No. 192), Eds. P. Whitelock and R.D. Cannon, (San Francisco: ASP), in press

Irwin, M.J., Bunclark, P.S., Bridgeland, M.T. & McMahon, R.G. 1990, MNRAS, 244, 16p

Irwin, M.J., Demers, S. & Kunkel, W.E. 1990, AJ, 99, 191

Jerjen, H., Freeman, K.C. & Binggeli, B. 1998, AJ, 116, 2885

Johnson, J.A., Bolte, M., Bond, H.E., Hesser, J.E., de Oliveira, C.M., Richer, H.B., Stetson, P.B. & VandenBerg, D.A. 1998, in New Views of the Magellanic Clouds (= IAU Symposium No. 190), Eds. Y.-H Chu et al., (San Francisco: ASP), in press

Jones, D.H. et al. 1996, ApJ, 466, 742

Kahn, F.D. & Woltjer, L. 1959, ApJ, 130, 705

Kenney, J.D.P. 1990, in The Interstellar Medium in Galaxies, Eds. H.A. Thronson and J.A. Shull, (Dordrecht: Kluwer), p. 151

King, I.R., Stanford, S.A. & Crane, P. 1995, AJ, 109, 164

Kinman, T.D. 1959, MNRAS, 119, 538

Kinman, T.D., Stryker, L.L., Hesser, J.E., Graham, J.A., Walker, A.R., Hazen,
M.L. & Nemec, J.M. 1991, PASP, 103, 1279

Kormendy, J. 1985, ApJ, 295, 73

Kormendy, J. 1988, ApJ, 325, 128

Kormendy, J. 1998, BAAS, 30, 1281

Kormendy, J. & McClure, R.D. 1993, AJ, 105, 1793

Krismer, M., Tully, R.B. & Gioia, I.M. 1995, AJ, 110, 1584

Kuijken, K. 1996, in Barred Galaxies (= ASP Conference Series Vol. 91), Eds. R.
Buta, D.A. Crocker and B.G. Elmegreen, (San Francisco:  ASP), p. 504

Lauer, T.R. et al. 1993, AJ, 106, 1436

Lauer, T.R. et al. 1996, ApJ, 471, L79

Lauer, T.R., Faber, S.M., Ajhar, E.A., Grillmair, C.J. & Scowen, P.A. 1998, AJ,
116, 2263

Lee, M.G., Freedman, W.L. & Madore, B.F. 1993c, AJ, 106, 964

Lee, Y.-W., Demarque, P. & Zinn, R. 1994, ApJ, 423, 248

Lerner, M.S., Sundin, M. & Thomasson, M. 1999, A&A, in press

Lindblad, B. 1927, MNRAS, 87, 553

Lo, K.Y., Sargent, W.L.W. & Young, K. 1993, AJ, 106, 507

Lynden-Bell, D. 1981, Observatory, 101, 111

Magorrian, J. & Tremaine, S. 1999, astro-ph/9902032

Majewski, S.R., Munn, J.A. & Hawley, S.L. 1996, ApJ, 459, L73

Marconi, G., Buonanno, R., Corsi, C.E. & Zinn, R. 1999, in The Stellar Content of Local Group Galaxies, (= IAU Symposium No. 192), Eds. P. Whitelock and R. Cannon, (San Francisco: ASP), in press

Martínez-Delgado, D., Gallart, C. & Aparicio, A. 1998, in Galactic Halos (= ASP Conf. Series Vol. 136), Ed. D. Zaritsky, (San Francisco: ASP), p. 81

Massey, P. & Armandroff, T.E. 1995, AJ, 109, 2470

Mateo, M. 1998, ARAA, 36, 435

Mateo, M., Fischer, P. & Krzeminski, W. 1995, AJ, 110, 2166

Mateo, M., Olszewski, E.W. & Morrison, H.L. 1999, ApJ, 508, L55

Mateo, M., Olszewski, E.W., Pryor, C., Welch, D.L. & Fischer, P. 1993, AJ, 105, 510

Mateo, M., Olszewski, E., Welch, D.L., Fisher, P. & Kunkel, W. 1991, AJ, 102, 914

Mathewson, D.S., Cleary, M.N. & Murray, J.D. 1974, in The Formation and Dynamics of Galaxies (= IAU Symposium No. 58), Ed. J.R. Shakeshift, (Dordrecht: Reidel), p. 367

Mathewson, D.S. & Ford, V.L. 1984, in Structure and Evolution of the Magellanic Clouds (= IAU Symposium No. 108), Eds. S. van den Bergh and K.S. de Boer, (Dordrecht: Reidel), p. 125

Mathewson, D.S., Ford, V.L. & Visvanathan, N. 1986, ApJ, 301, 664

Mathewson, D.S., Ford, V.L. & Visvanathan, N. 1988, ApJ, 333, 617

Mayall, N.U. 1935, PASP, 47, 317

McClure, R.D. & van den Bergh, S. 1968, AJ, 73, 313

McWilliam, A. & Rich, R.M. 1994, ApJS, 91, 749

Melia, F., Yusef-Zadeh, F. & Fatuzzo, M. 1998, ApJ, 508, 676

Merritt, D. 1999, Com. Ap. in press

Mighell, K.J. & Rich, R.M. 1996a, AJ, 111, 777

Mighell, K.J. & Rich, R.M. 1996b, in Formation of the Galactic Halo (= ASP

    Conference Series Vol. 92), Eds. H. Morrison and A. Sarajedini, (San

    Francisco:  ASP), p. 528

Minniti, D. 1996, ApJ, 459, 599

Montegriffo, P., Bellazzini, M., Ferraro, F.R., Martins, D., Sarajedini, A. & Fusi

    Pecci, F. 1998, MNRAS, 294, 315

Morgan, D.H. 1994, A&AS, 103, 235

Morgan, D.H. & Hatzidimitriou, D. 1995, A&AS, 113, 539

Morgan, W.W. 1959, AJ, 64, 432

Morgan, W.W., Whitford, A.E. & Code, A.D. 1953, ApJ, 119, 318

Morris, S.L. & van den Bergh, S. 1994, ApJ, 427, 696

Mould, J. & Aaronson, M. 1983, ApJ, 273, 530

Mould, J.R. & Kristian, J. 1986, ApJ, 305, 591

O'Connell, R.W. 1999, astro-ph/9903421

Olsen, K.A.G., Hodge, P.W., Mateo, M., Olszewski, E.W., Schommer, R.A.,

    Suntzeff, N.B. & Walker, A.R. 1998, MNRAS, 300, 665

Oort, J.H. 1927, Bull. Ast. Inst. Netherl., 3, 275

Oort, J.H. 1928, Bull. Ast. Inst. Netherl., 4, 269

Oort, J.H. 1966, Bull. Astr. Inst. Netherlands, 18, 421

Oort, J.H. 1970, A&A, 7, 381

Pascarelle, S.M., Windhorst, R.A. & Keel, W.C. 1998, AJ, 116, 2659

Pfenniger, D., Combes, F. & Martinet, L. 1994, A&A, 285, 79

Ponder, J.M. et al. 1998, AJ, 116, 2297

Preston, G.W., Beers, T.C. & Shectman, S.A. 1994, AJ, 108, 539

Pritchet, C. & van den Bergh, S. 1987, ApJ, 316, 517

Pritchet, C. & van den Bergh, S. 1988, ApJ, 331, 135

Pritchet, C.J. & van den Bergh, S. 1994, AJ, 107, 1730

Pritchet, C.J. & van den Bergh, S. 1999, AJ, in press

Reid, M.J. 1993, ARAA, 31, 345

Richer, H.B. et al. 1996, ApJ, 463, 602

Richer, M.G., McCall, M.L. & Stasińska, 1998, in Abundance Profiles:

      Diagnostic Tools for Galaxy History, (= ASP Conference Series Vol.

      147), Eds. D. Friedli et al., (San Francisco:  ASP), p. 254

Roberts, M.S. 1972, in External Galaxies and Quasi-stellar Objects (= IAU

      Symposium No. 44), Ed. D.S. Evans, (Reidel:  Dordrecht), p. 12

Rubin, V.C., Kumar, C.K. & Ford, W.K. 1972, ApJ, 177, 31

Sadler, E.M., Rich, R.M. & Terndrup, D.M. 1996, AJ, 112, 171

Saha, A., Hoessel, J.G. & Mossman, A.E. 1990, AJ, 100, 108

Sakai, S., Madore, B.F. & Freedman, W.L. 1999, ApJ, 511, 671

Sandage, A. 1986, ApJ, 307, 1

Sarajedini, A., Geisler, D., Harding, P. & Schommer, R. 1998, ApJ, 508, L37

Schechter, P.L. 1976, ApJ, 203, 297

Schmidt, M. 1954, Bull. Astr. Inst. Netherl., 13, 247

Schommer, R.A., Christian, C.A., Caldwell, N., Bothun, G.D. & Huchra, J. 1991,
    AJ, 101, 873

Schommer, R.A., Olszewski, E.W., Suntzeff, N.B. & Harris, H.C. 1992, AJ, 103,
    447

Schwarzschild, M. 1954, AJ, 59, 273

Scott, J.E., Friel, E.D. & Janes, K.A. 1995, AJ, 109, 1706

Searle, L. & Zinn, R. 1978, ApJ, 225, 357

Serabyn, E., Schupe, D. & Figer, D.F. 1998, Nature, 394, 448

Shapley, H. 1918a, ApJ, 48, 154

Shapley, H. 1918b, PASP, 30, 42

Shapley, H. 1938, Harvard Bull. #908, 1

Shapley, H. 1939, Proc. Nat. Acad. Sci. USA, 25, 565

Shapley, H. 1943, Galaxies (Philadelphia: Blakiston)

Shetrone, M.D., Bolte, M. & Stetson, P.B. 1998, AJ, 115, 1888

Spinrad, H., Taylor, B.J. & van den Bergh, S. 1969, AJ, 74, 525

Spitzer, L. 1969, ApJ, 158, L139

Stetson, P.B., Hesser, J.E. & Smecker-Hane, T.A. 1998, PASP, 110, 533

Suntzeff, N.B. 1992, in The Stellar Populations of Galaxies (= IAU Symposium No. 149), Eds. B. Barbuy and A. Renzini, (Dordrecht: Kluwer), p. 23

Suntzeff, N.B., Mateo, M., Terndrup, D.M., Olszewski, E.W., Geisler, D. & Weller, W. 1993, ApJ, 418, 208

Terndrup, D.M. 1988, AJ, 96, 884

Tinsley, B.M. & Spinrad, H. 1971, Ap&SS, 12, 118

Tolstoy, E. 1999, in The Stellar Content of the Local Group (= IAU Symposium No. 192), Eds. P.A. Whitelock and R. Cannon, (San Francisco: ASP), in Press

Toomre, A. 1977, in the Evolution of Galaxies and Stellar Populations, Eds. B.M. Tinsley and R.B. Larson (New Haven: Yale Univ. Press), p. 401

Tosi, M., Pulone, L., Marconi, G. & Bragaglia, A. 1998, MNRAS, in press

Tremaine, S. 1995, AJ, 110, 628

Tremaine, S.D., Ostriker, J.P. & Spitzer, L. 1975, ApJ, 196, 407

Trentham, N. 1998a, MNRAS, 294, 193

Trentham, N. 1998b, MNRAS, 295, 360

van den Bergh, S. 1964, ApJS, 9, 65

van den Bergh, S. 1969, ApJS, 19, 145

van den Bergh, S. 1972a, ApJ, 171, L31

van den Bergh, S. 1972b, ApJ, 178, L99

van den Bergh, S. 1974, ApJ, 191, 271

van den Bergh, S. 1976, ApJ, 203, 764

van den Bergh, S. 1981, A&AS, 46, 79

van den Bergh, S. 1986, AJ, 91, 271

van den Bergh, S. 1991, PASP, 103, 609

van den Bergh, S. 1993, ApJ, 411, 178

van den Bergh, S. 1994a, in The Local Group, Eds. A. Layden, R.C. Smith and J.
    Storm, (Garching: ESO), p. 3

van den Bergh, S. 1994b, AJ, 107, 1328

van den Bergh, S. 1996a, Observatory, 115, 103

van den Bergh, S. 1996b, ApJ, 471, L31

van den Bergh, S. 1996c, in Formation of the Galactic Halo…Inside and Out, Eds.
    H. Morrison and A. Sarajedini, (= ASP Conference Series Vol. 92) (San
    Francisco: ASP), p. 499

van den Bergh, S. 1996d, AJ, 108, 986

van den Bergh, S. 1998a, ApJ, 495, L79

van den Bergh, S. 1998b, ApJ, 505, L127

van den Bergh, S. 1998c, AJ, 116,1688

van den Bergh, S. 1998d, ApJ, 507, L39

van den Bergh, S. 1998e, Galaxy Morphology and Classification (Cambridge:
    University of Cambridge Press), p. 26

van den Bergh, S. 1998f, ApJ, 492, 41

van den Bergh, S. 1999a, in Stellar Content of the Local Group, (= IAU Symposium No. 192), Eds. P. Whitelock and R.D. Cannon, (San Francisco: ASP), in press

van den Bergh, S. 1999b, AJ, 117, xxx

van den Bergh, S. 2000, The Galaxies of the Local Group (Cambridge: Cambridge Univ. Press)

van den Bergh, S. & Henry, R.C. 1962, Publ. David Dunlap Obs., 2, 281

van den Bergh, S. & Lafontaine, A. 1984, AJ, 89, 1822

van den Bergh, S. & McClure, R.D. 1980, A&A, 80, 360

van den Bergh, S. & Pritchet, C.J. 1992, in The Stellar Populations of Galaxies (= IAU Symposium No. 149), Eds. B. Barbuy and A. Renzini, (Dordrecht: Reidel), p. 161

van der Kruit, P.C. 1989, in The Milky Way as a Galaxy, Eds. G. Gilmore, I.R. King and P.C. van der Kruit, (Sauverny: Geneva Observatory), p. 331

van der Marel, R.P., Cretton, N., de Zeeuw, P.T. & Rix, H.-W. 1998, ApJ, 493, 613

van Driel, W., Kraan-Korteweg, R.C., Binggeli, B. & Huchtmeier, W.K. 1998, A&AS, 127, 397

Vesperini, E. 1998, MNRAS, 299,1019

Vogt, S.S., Mateo, M., Olszewski, E.W. & Keane, M.J. 1995, AJ, 109, 151

Wakker, B.P. & van Woerden, H. 1997, ARAA, 35, 217

Wakker, B.P., van Woerden, H. & Gibson, B.K. 1999, in Stromlo Workshop on High-Velocity Clouds (= ASP Conference Series Vol. xxx), Eds. B.K. Gibson and M.E. Putman (San Francisco: ASP), in press

Walker, A.R. 1999, in Post Hipparcos Standard Candles, Eds. A. Heck and F. Caputo (Dordrecht: Kluwer), p. 125

Walterbos, R.A.M. & Braun, R. 1994, ApJ, 431, 156

Welch, G.A., Sage, L.J. & Mitchell, G.F. 1998, ApJ, 499, L209

Westerhout, G. 1954, Bull. Astr. Inst. Netherl., 13, 201

Westerlund, B.E. 1997, The Magellanic Clouds, (Cambridge: Cambridge Univ. Press)

White, S.D.M. 1979, MNRAS, 189, 831

Whitelock, P.A. & Cannon, R.D. 1999, Stellar Content of the Local Group, (= IAU Symposium No. 192, (San Francisco: ASP), in press

Wilson, A.G. 1955, PASP, 67, 27

Wirth, A. & Gallagher, J.S. 1984, ApJ, 282, 85

Woltjer, L. 1975, A&A, 42, 109

Worthey, G. 1998, PASP, 110, 888

Young, L.M. & Lo, K.Y. 1996, ApJ, 462, 203

Young, L.M. & Lo, K.Y. 1997, ApJ, 476, 127

Zaritsky, D. 1999, in The Galactic Halo, Eds. B.K. Gibbon, T.S. Axelrod and M.E. Putman (San Francisco: ASP), in press

Zaritsky, D., Olszewski, E.W., Schommer, R.A., Peterson, R.C. & Aaronson, M.

   1989, ApJ, 345, 759

Zinn, R. 1985, ApJ, 293, 424

Zwicky, F. 1942, Phys, Rev: II, 61, 489

**Acknowledgements**

I should like to take this opportunity to extend my heartfelt thanks to Stéphane Courteau, Eva Grebel, Jim Hesser, Mario Mateo, and Chris Pritchet for their input into this paper.

**Table 1** - Data on Local Group Members

| Name | Alias | α | (J2000) | δ | DDO Type | M_V | V_r | D(Mpc) | D_LG (Mpc) |
|------|-------|---|---------|---|----------|-----|-----|--------|------------|
| WLM | DDO 221 | 00ʰ 01ᵐ 57.8 | -15 27′ 51 | | Ir IV-V | -14.4 | -120 | 0.93 | 0.79 |
| IC 10 | UGC 192 | 00 20 24.5 | 59 17 30 | | Ir IV: | -16.3 | -344 | 0.66 | 0.27 |
| NGC 147 | UGC 326 | 00 33 11.6 | 48 30 28 | | Sph | -15.1 | -193 | 0.66 | 0.22 |
| And III | A0032+36. | 00 35 17.0 | 36 30 30 | | dSph | -10.2 | ... | 0.76 | 0.31 |
| NGC 185 | UGC 396 | 00 38 58.0 | 48 20 18 | | Sph | -15.6 | -202 | 0.66 | 0.22 |
| NGC 205 | ... | 00 40 22.5 | 41 41 11 | | Sph | -16.4 | -244 | 0.76 | 0.31 |
| M 32 | NGC 221 | 00 42 41.9 | 40 51 55 | | E2 | -16.5 | -205 | 0.76 | 0.31 |
| M 31 | NGC 224 | 00 42 44.2 | 41 16 09 | | Sb I-II | -21.2† | -301 | 0.76 | 0.30 |
| And I | A0043+37 | 00 45 43 | 38 00 24 | | dSph | -11.8 | ... | 0.81 | 0.36 |
| SMC | ... | 00 52 36 | -72 48 00 | | Ir IV/IV-V | -17.1 | 148 | 0.06 | 0.48 |
| Sculptor | ... | 01 00 04.3 | -33 42 51 | | dSph | - 9.8 | 110 | 0.09 | 0.44 |
| Pisces | LGS 3 | 01 03 56.5 | 21 53 41 | | dIr/dSph | -10.4 | -286 | 0.81 | 0.42 |
| IC 1613 | ... | 01 04 47.3 | 02 08 14 | | Ir V | -15.3 | -232 | 0.72 | 0.47 |
| And V | ... | 01 10 17.1 | 47 37 41 | | dSph | -10.2 | ... | 0.81 | 0.37 |
| And II | ... | 01 16 27 | 33 25 42 | | dSph | -11.8 | ... | 0.70 | 0.26 |
| M 33 | NGC 598 | 01 33 50.9 | 30 29 37 | | Sc II-III | -18.9 | -181 | 0.79 | 0.37 |
| Phoenix | ... | 01 51 03.3 | -44 27 11 | | dIr/dSph | - 9.8 | ... | 0.40 | 0.59 |
| Fornax | ... | 02 39 53.1 | -34 30 16 | | dSph | -13.1 | 53 | 0.14 | 0.45 |
| LMC | ... | 05 19 36 | -69 27 06 | | Ir III-IV | -18.5 | 275 | 0.05 | 0.48 |
| Carina | ... | 06 41 36.7 | -50 57 58 | | dSph | - 9.4 | 223 | 0.10 | 0.51 |
| Leo A | DDO 69 | 09 59 23.0 | 30 44 44 | | Ir V | -11.5 | 24 | 0.69 | 0.88 |
| Leo I | Regulus | 10 08 26.7 | 12 18 29 | | dSph | -11.9 | 287 | 0.25 | 0.61 |
| Sextans | ... | 10 13 02.9 | -01 36 52 | | dSph | - 9.5 | 226 | 0.09 | 0.51 |
| Leo II | DDO 93 | 11 13 27.4 | 22 09 40 | | dSph | -10.1 | 76 | 0.21 | 0.57 |
| Ursa Min. | DDO 199 | 15 08 49.2 | 67 06 38 | | dSph | - 8.9 | -247 | 0.06 | 0.43 |
| Draco | DDO 208 | 17 20 18.6 | 57 55 06 | | dSph | - 8.6 | -293 | 0.08 | 0.43 |
| Milky Way | Galaxy | 17 45 39.9 | -29 00 28 | | S(B)bc I-II | -20.9 | 16 | 0.01 | 0.46 |
| Sagittarius | ... | 18 55 04.3 | -30 28 42 | | dSph(t) | -13.8:: | 142 | 0.03 | 0.46 |
| SagDIG * | ... | 19 29 58.9 | -17 40 41 | | Ir V | -10.7: | - 79 | 1.40: | 1.48 |
| NGC 6822 | ... | 19 44 56.0 | -14 48 06 | | Ir IV-V | -16.0 | - 56 | 0.50 | 0.67 |
| Aquarius* | DDO 210 | 20 46 53 | -12 50 58 | | V | -11.3 | -131 | 1.02 | 1.02 |
| Tucana* | ... | 22 41 48.9 | -64 25 21 | | dSph | - 9.6 | ... | 0.87 | 1.10 |
| Cassiopeia | And VII | 23 26 31 | 50 41 31 | | dSph | - 9.5 | ... | 0.69 | 0.29 |
| Pegasus | DDO 216 | 23 28 34 | 14 44 48 | | Ir V | -12.3 | -182 | 0.76 | 0.44 |
| Pegasus II | And VI | 23 51 39.0 | 24 35 42 | | dSph | -10.6 | ... | 0.83 | 0.43 |

\*      Possible LG members
†      Quoted luminosity would increase if internal absorption corrections were made for this nearby edge-on galaxy.

**Table 2** - Sub-Structure in the Local Group[1]

M 31
M 32
NGC 205

}

NGC 147
NGC 185

}

M 33

}            Andromeda subgroup

IC 10

NGC 6822

IC 1613

WLM

LMC
SMC

}

Galaxy

}            Galaxy subgroup

[1]      Only galaxies with $M_V$ < -14.0 listed

**Table 3** - Luminosity Distributions in Galaxy and M 31 Subgroups

| Luminosity range | M 31 subgroup | Galaxy subgroup |
|---|---|---|
| $-15.0 < M_V < -10.0$ | 6 | 4 |
| $-10.0 < M_V < -5.0$ | 1 | 5 |

**Table 4 -** Local Group properties

| | | |
|---|---|---|
| Radius of zero-velocity surface | $R_o = 1.18 \pm 0.16$ Mpc | (1) |
| Half-mass radius | $R_h \approx 350$ kpc | (2) |
| Total LG luminosity | $L_V = 4.2 \times 10^{10}$ $L_\odot$ | (3) |
| Radial velocity dispersion | $\sigma_r = 61 \pm 8$ km s$^{-1}$ | (4) |
| Total LG mass | $M = (2.3 \pm 0.6) \times 10^{12}$ $M_\odot$ | (5) |
| LG mass-to-light ratio | $M/L_V = 44 \pm 14$ | (6,9) |
| Mass of Andromeda subgroup | $M(A) = 1.15$-$1.5 \times 10^{12}$ $M_\odot$ | (7) |
| Luminosity of Andromeda subgroup | $L_V = 3.0 \times 10^{10}$ $L_\odot$ | (5) |
| Andromeda subgroup M/L$_V$ | $M/L_V = 38$ - $50$ | (5,6,9) |
| Mass of Milky Way subgroup | $M(G) = 0.46$-$1.25 \times 10^{12}$ $M_\odot$ | (8) |
| Luminosity of Milky Way subgroup | $L_V = 1.1 \times 10^{10}$ $L_\odot$ | (9) |
| Galactic subgroup M/L$_V$ | $M/L_V = 42$ -$144$ | (6) |

Notes and references:

(1)     From Courteau & van den Bergh (1999).  The quoted error assumes (a) that the zero-velocity surface is spherical, (b) that the Local Group age is $T_o = 14 \pm 2$ Gyr, and (c) that its mass is $(2.3 \pm 0.6) \times 10^{12}$ $M_\odot$ .

(2)     Courteau & van den Bergh find that the radius containing half of all LG members is $\approx$ 450 kpc.

(3)     Errors mainly due to poorly determined absorption in M 31, and uncertainty of the $M_V$ value for the Milky Way system.  Quoted luminosity corresponds to $M_V = -22.0$.

(4)     From solar motion solution of Courteau & van den Bergh (1999).

(5)     From Courteau & van den Bergh (1999).

(6)     M/L$_V$ in solar units.

(7)     The lower value is derived from the virial theorem, and the upper value from the projected mass method (Bahcall & Tremaine 1981, Heisler, Tremaine & Bahcall 1985).

(8)     From Zaritsky et al. (1989) and Zaritsky (1998).  The lower limit assumes that Leo I is not a satellite of the Galaxy.

(9)     Luminosity from data in Table 1.  No internal absorption corrections were applied to M 31 luminosity.

**Table 5** - Comparison between M 31 and Galaxy

|  | Andromeda Galaxy | Milky Way system |
|---|---|---|
| Type | Sb  I-II | S(B)bc  I-II: |
| $L_V$[1] | $2.6 \times 10^{10} L_\odot$ | $2: \times 10^{10} L_\odot$ |
| D | $760 \pm 35$ kpc | $8.5 \pm 0.5$ kpc |
| $M_{H\,I}$ | $5.8 \times 10^9 M_\odot$ | $4 \times 10^9 M_\odot$ |
| M(nucleus) | $7 \times 10^7 M_\odot$ | $2.5 \times 10^6 M_\odot$ |
| Bulge light[1] | 25% | 12% |
| $R_e$ (bulge) | 2.4 kpc | 2.5 kpc |
| Bulge dispersion | $\sigma = 155$ km s$^{-1}$ | $\sigma = 130$ km s$^{-1}$ |
| Rotational velocity[1] | $\sim 260$ km s$^{-1}$ | $\sim 220$ km s$^{-1}$ |
| No. globulars | $400 \pm 55$ | $160 \pm 20$ |

[1]    van der Kruit (1989).  No internal absorption corrections applied.